\documentclass[preprint2]{aastex}

\shorttitle{Radiation from collisions of compact 
objects with jets}
\shortauthors{Bednarek \& Banasi\'nski}

\begin{document}


\title{Non-thermal radiation from collisions of compact objects with intermediate scale jets in active galaxies}


\author{W. Bednarek \& P. Banasi\'nski}
\affil{Department of Astrophysics, The University of Lodz, 90-236 Lodz, ul. Pomorska 149/153, Poland}

\email{bednar@uni.lodz.pl}



\begin{abstract}
Massive black holes in active galaxies are immersed in huge concentrations of late type stars in the galactic bulges and also early type massive stars in the nuclear stellar clusters which are additionally surrounded 
by quasi-spherical several kpc scale halos containing from a few hundred up to several thousand globular
clusters (GCs). It is expected that significant numbers of red giant stars, massive stars and also GCs
can move through the jet expelled from the central engine of active galaxy. We consider collisions of stars from the galactic bulge, nuclear cluster and globular clusters with the jet plasma. As a result of such collisions, multiple shocks are expected to appear in the jet around these compact objects. Therefore, the plasma in the  kpc scale jet can be significantly disturbed. We show that particles can be accelerated on these shocks up to the multi-TeV energies. TeV leptons emit synchrotron radiation, extending up to the X-ray energies, and also comptonize radiation produced in a stellar cluster and also the Microwave Background Radiation to TeV $\gamma$-ray energies. 
We show that such non-thermal radiation is likely to be detectable from the intermediate scale jets of 
the nearby active galaxies for reasonable number of stars and GCs immersed within the jet. 
As an example, we calculate the expected non-thermal emission in the X-ray and gamma-ray energies from 
the nearby radio galaxy Cen A from which a steady gamma-ray emission with the complex spectrum has been 
recently reported by the {\it Fermi} and the HESS Observatories. 

\end{abstract}





%
%
\section{Introduction}

The non-thermal X-ray emission from large scale jets in nearby radio galaxies of the Fanaroff-Riley class I
is expected to be common feature of such type of objects starting from its discovery in two close galaxies Cen A or M 87 (Feigelson et al.~1981, Biretta et al.~1991).
This X-ray emission smoothly extends up to the radio energy range supporting the idea on its common origin in the synchrotron process of TeV electrons (Hardcastle et al.~2001). TeV electrons are expected to meet strong soft radiation field, produced either in the synchrotron process or by the stars in the galactic bulge and also the Microwave Background Radiation (MBR), fro its comptonization to  TeV energies (e.g. Stawarz et al.~2003,
Hardcastle \& Croston~2011). Observations of persistent TeV $\gamma$-ray emission from such sources should provide important constraints on the high energy particles in the kpc scale jets.  
However, since such electrons has to lose energy on a short time scale, they require an {\it in situ} acceleration process. Therefore, efficient acceleration of electrons should occur not only within the inner jet but also in the kpc scale jets. In fact, the observed X-ray knots in the intermediate scale jets seems to provide conditions for electron acceleration to TeV energies. These might correspond to shocks formed in collisions of winds from compact objects with the jet plasma (e.g. Hardcastle et al.~2003).

A jet, launched from the vicinity of the central super-massive black hole (SMBH), has to pass through the stellar bulge and globular cluster (GC) halo around central engine. Since the  number of stars in the central stellar cluster is expected to be huge, many of them have to be immersed in the jet plasma. In fact, it has been argued that as a result of the star-jet collisions, stellar winds can provide large amount of barion matter into the jet (e.g. Komissarov~1994,  Bowman et al.~1996, Laing \& Bridle~2002, Hubbard \& Blackman~2006). Shocks formed in collisions of stellar winds with the plasma in the inner jet can be also responsible for the acceleration of electrons which, comptonizing stellar radiation, might be responsible for the production of $\gamma$-ray flares 
(e.g. Bednarek \& Protheroe 1997, Barkov et al.~2010, Bosch-Ramon et al.~2012, Araudo et al.~2013, Wykes et al.~2014). Also collisions of relativistic jet plasma with dense clouds can turn to the $\gamma$-ray production (e.g. Blandford \& K\"onigl~1979, Dar \& Laor 1997, Beall \& Bednarek~1999, Purmohammad \& Samimi~2001, Araudo et al.~2010,
Barkov et al.~2012). These clouds might be fragments of the close to the jet supernova explosion Blandford \& K\"onigl~(1979). Large number of objects, moving in the galactic bulge, are expected to be immersed within the kpc 
scale jets.  

Galaxies of different types are surrounded by spherical halos of GCs with the radii
of several kpcs. GCs are expected to produce winds which are the products of the mixture of the relativistic winds from a large population of millisecond pulsars (MSPs) within the GC and slow
barionic winds from the red giant stellar population within GCs (Bednarek \& Sobczak~2014). The mixed  millisecond pulsar and stellar winds, expelled from the GC, interact with a relatively rare relativistic intermediate scale jet. As a result, a double shock structure forms around GC. We argue that particles can be accelerated on such shocks to multi-TeV energies. They produce non-thermal radiation by interacting with the magnetic field in the jet (the synchrotron radiation) and with the soft radiation from the galactic bulge at the sub-kpc distance scale and also with the MBR at the a kpc distance scale via inverse Compton process.

In this paper we explore the hypothesis that relativistic electrons appear in the intermediate scale jet 
as a result of their acceleration on the shocks formed in collisions of many compact objects 
(red giants, massive stars or GCs) with the jet. It is assumed that at the kpc scale distances jets are already non-relativistic and the Doppler boosting effects are small.
Under this assumption, we calculate the synchrotron and inverse Compton $\gamma$-ray emission produced in the intermediate jet for different parameters of such scenario.
As an example, we show that the high energy TeV $\gamma$-ray component in the Cen A spectrum, observed by the HESS Collaboration (Aharonian et al.~2009, Sahakyan et al.~2013), might originate within the intermediate scale jet as a result of acceleration of electrons in collisions of many compact objects with the jet plasma.

\section{Stars and their clusters around AGN jets}

Active galaxies are complicated systems surrounded at larger scale by a few important elements. The central
SMBH, with the mass in the range $\sim 10^6-10^{10}$ M$_\odot$, is surrounded by the bulge, which has quasi-spherical shape, containing late type stars within the radius of the order of $\sim$kpc. The mass of the bulge is related to the mass of the SMBH and its luminosity is of the order of a few tens percent of the whole galaxy. Many SMBHs are surrounded by central clusters of luminous and massive stars. Such clusters have usually extended shape with the main axis in the galactic plane. Its  mass and dimension is supposed to be determined by the destruction process of another galaxy by the massive galaxy harboring SMBH. Finally, active galaxy is surrounded by a halo of GCs within the distance scale of several kpc. The number of GCs is related to the mass of the parent galaxy and the mass of the SMBH. The parameters of these stellar clusters around nucleus of active galaxy are discussed below.

Nuclear clusters of massive stars are observed not only around SMBHs in massive active galaxies but also in the center of our Galaxy. In the case of our Galaxy such cluster is relatively small. 
It has a half mass radius of $\sim$ pc and the mass of $\sim 2.5\times 10^7$ M$_\odot$ (e.g. Sch\"odel et al.~2014)
In the case of the nearby active radio galaxy Cen A, the massive stellar cluster has been formed around 50-100 Myrs ago as a result of a merger process with a gas reach galaxy. The observed star formation region is centered on the Cen A nucleus. It has the visual dimensions of $\sim 8\times$3 kpc$^2$ (Neff et al.~2015). The current star formation rate has been estimated on 2 M$_\odot$yr$^{-1}$ which will result in production of $(6-12)\times 10^7$ M$_\odot$ of young stars (Wykes et al.~2014). Assuming the standard initial mass function of stars, we estimate the total number of stars with masses above 20 M$_\odot$ in the nuclear cluster in Cen A on $\sim 3\times 10^5$. A small part of these stars can pass through the jet expelled from the super-massive black hole. The rough estimate of the number of massive O and WR type stars, contained within the 1.5 kpc from the base of the jet with the opening angle of the order of 0.1 rad for the cluster with dimension mentioned above, is of the order of a hundred. This 
simple estimate shows that many massive stars can interact with the jet plasma. 

Central regions of galaxies are also surrounded by the quasi-spherical concentrations of late type stars (galactic bulges) with typical dimensions of the order of $\sim$1 kpc and luminosity of several percent of the total luminosity of the parent galaxy. Interestingly, the masses of bulges are related to masses of the central black holes (for review on scaling formulas see e.g. Czerny \& Niko\l ajuk~2010). In the case of Cen A, the number of bulge stars within the jet has been estimated on $\sim 8\times 10^8$ (Wykes et al.~2014). If only a small number of these stars are in the red giant phase, of the order of $\sim$10$^{-3}$, then the total number of red giants in Cen A jet can be as large as $\sim 10^6$.   

As we mentioned in the Introduction also GCs are expected to enter the jet from time to time at the kpc distances from the base of the jet. In fact the number of observed (or estimated) GCs around galaxies seems large enough to guarantee encounters with the jet. For example, in the case of our Galaxy more than 150 has been discovered (Harris~2003).
Larger galaxies are expected to have more GCs. The Andromeda galaxy is expected to be surrounded by a halo of $\sim$500 GCs (Barmby \& Huchra~2001) but the giant radio galaxy, M87, contains already $\sim$13000 GCs 
(McLaughlin et al.~1994). It is shown that in active galaxies, containing super-massive black holes (SMBHs), 
the number of GCs in the galactic halo is correlated with the mass of the central black hole
(e.g. Spitler \& Forbes~2009, Burkert \& Tremaine~2010, Harris et al.~2014). For example, in the radio galaxy Cen A the number of GCs is estimated on $\sim 1550$ (G\"ultekin et al.~2009).

\begin{figure*}
\vskip 6.truecm
\includegraphics{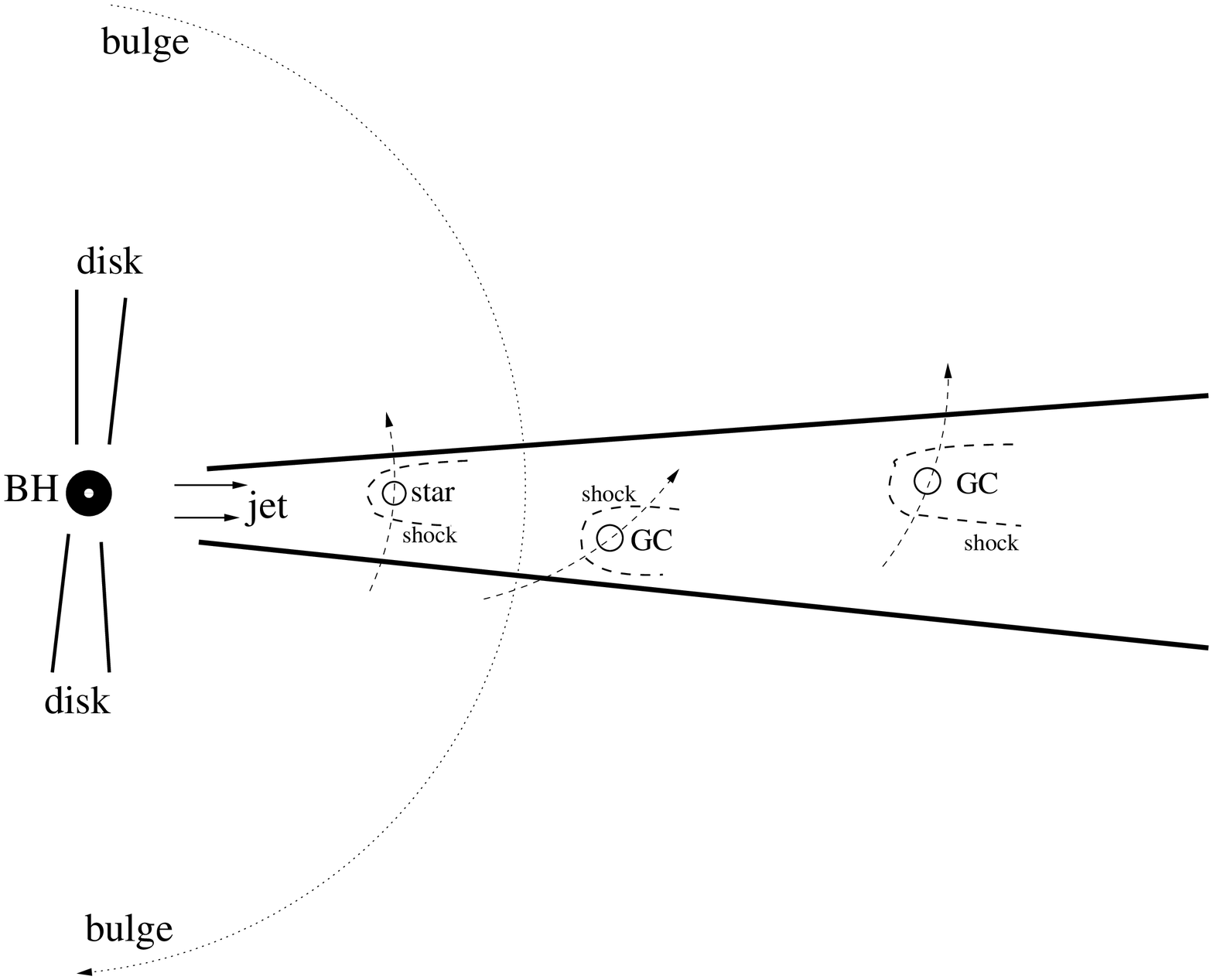}
\includegraphics{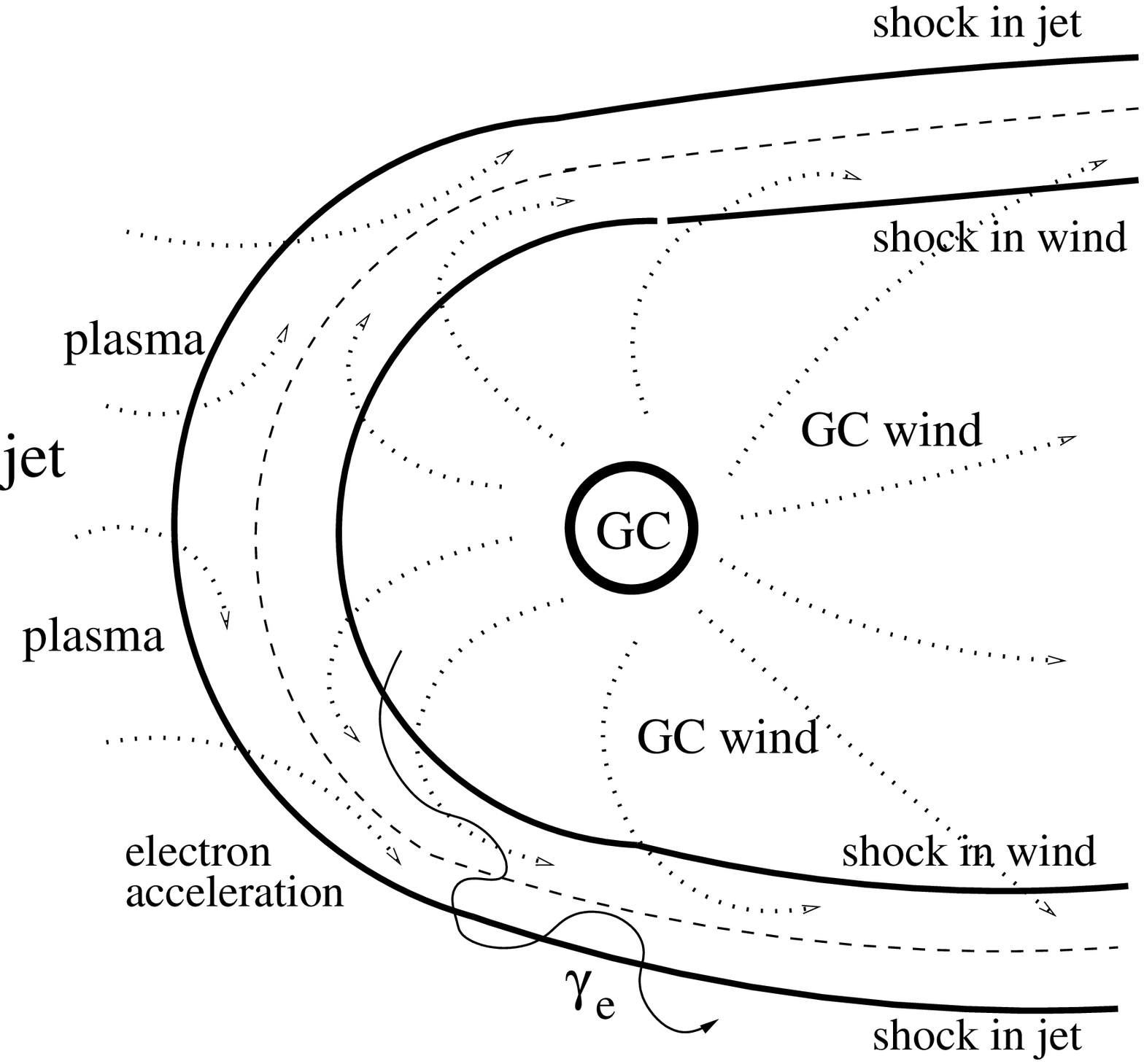}
\caption{Schematic representation (not to scale) of the interaction of the winds from compact objects (red giants, massive stars,  globular clusters) with the plasma of a relativistic jet in active galactic nuclei. On the left: large number of stars and GCs from halos around the central engine on the distance scale up to $\sim$kpc and several kpcs, respectively. Some of these objects enter the jet expelled by the central super-massive black hole. On the right:
The example case of collision of a GC with a jet. GCs are expected to contain up to a few hundred of millisecond pulsars (MSPs) and a few hundred thousand of normal stars, many of them in the red giant phase. The MSP winds and red giant winds mix efficiently within the cluster producing mixed pulsar-stellar wind from the GC.
This GC wind interacts with the jet plasma. As a result, a double shock structure is formed. Leptons can be accelerated at the shock up to multi-TeV energies. These electrons lose energy on the synchrotron process and on the IC scattering of the soft radiation produced by stars in the bulge (at kpc distance scales) and on the Microwave Background Radiation during their propagation along the kpc scale jet.}
\label{fig1}
\end{figure*}
\section{Interaction of compact objects with AGN jets}
 
We consider active galaxy which has a jet propagating from the SMBH. The jet has a conical structure on the intermediate scale. The jet opening angle at a kpc distance scale is expected to be of the order of $\theta\sim 0.1\theta_{-1}$ rad, see e.g. observations of the nearby active galaxy Cen A (e.g. Clarke et al.~1992, Hardcastle \& Croston~2011). Then, the solid angle subtended by the jet is, $\Delta\Omega = 2.5\times 10^{-3}\theta_{-1}^2$. 
Since the number of compact objects surrounding central engine can be very large (see above), 
many of the late type stars, massive stars, and GCs can pass through the jet (see Fig.1 on the left).
As we mentioned above the number of these compact objects is shown to be related to the mass of the central black hole. Therefore, it is expected that very massive black holes, which are on average more luminous, are also surrounded by a larger number of stars and stellar clusters. For example,  
we estimate the number of GCs within the jets of two well known nearby active galaxies, Cen A and M87, on $\sim 10$ and $\sim$70 GCs for the jet opening angle equal to 0.15 rad.  

The power of jets in the nearby radio galaxies are typically of the order of $\sim$10$^{43}$ erg s$^{-1}$ (e.g. 
in Cen A, Wykes et al.~2013).
The ram pressure of the jet plasma can be estimated from (e.g. Bednarek \& Protheroe~1997), 
\begin{eqnarray}
P_{\rm j} = {{L_{\rm j}}\over{\pi c\theta^2l^2}}\approx 1.1\times 10^{-9}L_{43}\theta_{-1}^{-2}l_3^{-2}~~~{\rm {{erg}\over{cm^{3}}}},
\label{eq1}
\end{eqnarray} 
\noindent
where the power of the jet is $L_{\rm j} = 10^{43}L_{43}$ erg s$^{-1}$, the distance from the base of the jet is $l = 1l_3$ kpc and $c$ is the velocity of light. Compact objects, mentioned above, are also surrounded by winds which pressure can be balanced by the pressure of the jet plasma. 
As a result, a double shock structure is expected to appear around the compact objects within the jet as already discussed in the case of different types of stars. In the case of GCs, hadronic winds from significant population of red giant stars mix efficiently with the winds from many millisecond pulsars (MSPs) within the GC. As a result, the wind emanates from the GC with the power determined by the energy output from the MSP population within the GC. This mixture of red giant/MSP winds collide with the jet's plasma as shown schematically in Fig. 1 (on the right).
We argue that particles can be accelerated at the shocks from the jet 
site to multi-TeV energies. They produce non-thermal radiation by interacting with the magnetic field in the jet (the synchrotron radiation) and with the soft radiation from the galactic bulge at the kpc distance scale and also with the Microwave Background Radiation (MBR) at the distance scale up to a few tens of kpcs (the inverse Compton radiation). Below we consider separately the case of collision of a star with the jet and a GC with the jet since these types of objects differ significantly in the parameters describing their winds.

\subsection{Stars in the galactic bulge and nuclear stellar cluster}

As we argue above two different classes of stars are expected to surround the central engine of active galaxy
at distances up to the order of a kpc. They are massive stars from the nuclear stellar cluster and late type stars (red giants) from the galactic bulge. In fact, collisions of the winds from massive stars with the inner jet plasma in active galaxies, as a possible scenario for acceleration of particles and their radiation,  have been  already discussed in Bednarek \& Protheroe~(1997). More recently, distraction process of red giant stars as a result of collisions with the inner jets has been also considered as the mechanism for $\gamma$-ray flares    
(e.g. Barkov et al.~2010 and Bosch-Ramon et al.~2012). Such destruction process is possible only in the case of
very powerful jets and at a few pc scale distances from the base of the jet. 
At large distances, where the pressure of the jet plasma is weak, the red giant star winds are strong enough
to stop the jet plasma above the red giant surface. As a result, the scenario for collisions of red giant stars becomes similar to the scenario of collisions of massive stars with jets on the intermediate scale distances from
the base of the jet.  Recently, collision of different types of stars with the intermediate scale jets, as the mechanism for the non-thermal synchrotron emission from jets, has been discussed by Wykes et al.~(2014) with the application to Cen A.

First, we consider collisions of stars with parameters of the red giant stars. The scenario for collisions of massive stars will be in principle the same. We will comment on possible differences at the end of this section.
We assume that a small part of the bulge stars, of the order of $\sim 10^{-3}$, are in the phase of the red giant. As an example, we consider red giants with following typical parameters, the radius $R_{\rm RG} = 5\times 10^{12}$ cm, the surface temperature $T_{\rm RG} = 3700$ K, the luminosity $L_{\rm RG} = 10^3$ L$_\odot$, the mass loss rate $\dot{M}_{\rm w} = 10^{-7}M_{-7}$ M$_\odot$ yr$^{-1}$, and the wind velocity $v_{\rm w} = 3\times 10^6v_3$ cm s$^{-1}$.
As discussed in Sect.~2, a large number of red giant stars, of the order of $N_{\rm RG} = 10^{6} N_6$, is expected to be immersed in the intermediate jet of active galaxy on the distance scale of $\sim 1$ kpc (see estimates for the numbers of bulge stars within the jet in Wykes et al.~2014). 

These red giant stars produce stellar winds which pressure can be estimated from,
\begin{eqnarray}
P_{\rm w}^\star = {{\dot{M}_{\rm w}v_{\rm w}}\over{4\pi R^2}}\approx 
1.6\times 10^{-13}M_{-7}v_3R_1^{-2}~~~{\rm {{erg}\over{cm^{3}}}},
\label{eq2}
\end{eqnarray}
\noindent
where the distance from the star is $R = 1R_1$ pc.
The stellar wind pressure balances the pressure of the jet plasma at the distance from the stellar surface at
(Bednarek \& Protheroe~1997),
\begin{eqnarray}
R_{\rm sh}^\star\approx 3.6\times 10^{16}(M_{-7}v_3)^{1/2}\theta_{-1}l_3/L_{43}^{1/2}~~~{\rm cm}.
\label{eq3}
\end{eqnarray}
\noindent
For small distances from the base of the jet, the pressure of the jet might not be balanced by the pressure of 
the wind of the red giant above its stellar surface. In fact, the shock is above the red giant surface for the stars which enter the jet at the distances from its base which are larger than,
\begin{eqnarray}
l\approx 0.15 L_{43}^{1/2}/[\theta_{-1}(M_{-7}v_3)^{1/2}]~~~{\rm pc}.
\label{eq4}
\end{eqnarray}
\noindent
For the jet powers as observed in nearby radio galaxies (e.g. Cen A), $L = 10^{43}$ erg s$^{-1}$, this distance scale is only of the order of $l\sim 0.1$ pc. It is much smaller than the radius of the galactic bulge. Therefore, most of the red giants entering the jet in Cen A from its galactic bulge will be surrounded by the shocks clearly at some distance from stellar surface. The atmospheres of these red giants will not suffer direct collisions with the jet plasma as considered by Araudo et al.~(2010) and Barkov et al.~(2010). Instead, the jet plasma will be stopped at some distance from the red giant surface creating a shock as considered by Bednarek \& Protheroe~(1997) for the case of massive stars entering the inner jet in active galaxies.

The maximum power, that can be extracted from the jet by such a shock structure around a single star, is estimated on,
\begin{eqnarray}
L_{\rm sh}^\star = L_{\rm j}({{R_{\rm sh}^\star}\over{\theta l}})^2\approx 1.45\times 10^{35}M_{-7}v_3~~~{\rm {{erg}\over{s}}}.
\label{eq5}
\end{eqnarray}  
Note that this power depends only on the parameters of the red giants star. Since the total number of red giants in the jet can be quite large, a significant energy of the jet can be extracted and eventually transferred to relativistic particles in the acceleration mechanism operating at the collision region of the jet and stellar winds. 

The parameters of the massive stars, which can also enter the jet, are clearly much more extreme than expected for the red giants. For example, the terminal velocities of the winds of massive O type stars can be $v_{\rm w}\sim (1 - 3)\times 10^3$ km s$^{-1}$ 
and the mass loss rates $\dot{M}_{\rm w}\sim 10^{-6}$ M$_\odot$ yr$^{-1}$ and of the WR type stars $\sim (1 - 5)\times 10^3$ km s$^{-1}$ 
and the mass loss rates $\sim (0.8 - 8)\times 10^{-5}$ M$_\odot$ yr$^{-1}$, respectively (Lang~1992). Stars in different phases of evolution can have very large mass loss rates of the order of
$10^{-5} - 10^{-2}$ M$_\odot$ yr$^{-1}$, e.g. Asymptotic Giant Branch stars or Luminous Blue Variable stars (see Table~1 in Wykes et al.~2014). These massive stars, although more rare than
red giants, will produce shocks in jets which are able to extract $\sim$3-4 orders of magnitude more energy from the jet plasma than a single red giant.

\subsection{Globular clusters in the galactic halo}

As described in Sect.~2, GCs are expected to create a halo on several kpc distance scale around the central engine of active galaxy. A substantial number should be submerged in the jet with the opening angle of the order of $\sim$0.1-0.15 rad. GCs are composed from old low mass stars which significant number,
on average about $\sim$100, is expected to be in the red giant phase. These red giants are expected to be the main contributors of the mass into the GC through their stellar winds. In fact, the mass loss rate of isolated 
red giant is expected to be in the range between $10^{-9}$ M$_\odot$ yr$^{-1}$ to 
$3\times 10^{-7}$ M$_\odot$ yr$^{-1}$ (Boyer et al.~2008, Meszaros et al.~2009). The mass loss rate in the range of  $10^{-7}$ M$_\odot$ yr$^{-1}$ to $3\times 10^{-5}$ M$_\odot$ yr$^{-1}$ is estimated from 100 red giants in GC. It is expected that the single solar mass star should lose about 
$\sim 0.3$ M$_\odot$ during its evolution path to the White Dwarf (e.g. see discussion in Heyl et al.~2015). 
Simple estimates of the mass loss rates in a single GC, with the total number of stars equal to $10^6$, gives us the average mass loss rate during 10 Gyr equal to $\sim 3\times 10^{-5}$ 
M$_\odot$ yr$^{-1}$, consistently with the above estimated value. In our example calculations, we use the 
mass loss rates of stars in GCs close to this value. On the other hand, GCs contain many millisecond pulsars which has been accelerated to short periods due to the transfer of momentum from the stellar companions. These MSPs have typical surface magnetic fields of the order of $\sim$10$^8$ G and rotational periods of the order of $\sim$3 ms. The number of such pulsars in a single GC can be of the order of $\sim$100, as estimated from their observed $\gamma$-ray luminosity (Abdo et al.~2010c). These MSPs lose rotational energy in 
the form of the relativistic pulsar winds that should mix efficiently with slowly moving winds from the red giants (Bednarek \& Sobczak~2014). As a result, a barion loaded pulsar wind slows down to the sub-relativistic velocities.
The velocity of the mixed pulsar-stellar wind is estimated to be,
\begin{eqnarray}
v_{\rm w}^{\rm GC} =  \sqrt{{{2L_{\rm MSP}}\over{{\dot M}_{\rm GC}}}}\approx 
5.6\times 10^7 ({{L_{36}}\over{M_{-5}}})^{1/2}~~~{\rm {{cm}\over{s}}},
\label{eq6}
\end{eqnarray} 
where the mass loss rate of stars within the GC 
scales with ${\dot M}_{GC} = 3\times 10^{-5}M_{-5}$ M$_\odot$ yr$^{-1}$ and $L_{\rm MSP} = 3\times 10^{36}L_{36}$ erg s$^{-1}$ is the rotational energy loss rate of all millisecond pulsars within the GC.

The pressure of the GC wind is estimated from (Bednarek \& Sobczak~2014), 
\begin{eqnarray}
P_{\rm w}^{\rm GC} = {{\dot{M}_{\rm GC}v_{\rm w}^{GC}}\over{4\pi R^2}}\approx 
{{9.5\times 10^{-10}(L_{36}M_{-5})^{1/2}}\over{R_1^2}}~~~{\rm {{erg}\over{cm^{3}}}}.
\label{eq7}
\end{eqnarray} 
\noindent
The location of this shock in the jet around the GC, $R_{\rm sh}$, can be estimated by comparing the pressure of the wind from the GC with the ram pressure
of the jet plasma,
\begin{eqnarray}
R_{\rm sh}^{\rm GC}\approx 0.93 (L_{36}M_{-5})^{1/4}\theta_{-1}l_3/L_{43}^{1/2} ~~~{\rm pc}.
\label{eq8}
\end{eqnarray} 
\noindent
The radius of the shock should be larger than the core radius of the GC. Note that, most of GCs ($\sim 90\%$)
in our Galaxy have the core radius below 1 pc (Harris~1996). Therefore, for the parameters considered in 
this paper we expect that this condition is usually fulfilled for the GCs at the distance larger than $\sim$1 kpc from the central engine.

The shock structure, formed in collision of the jet with GC wind, is expected to have a double structure with different conditions on both sites (see Fig.~1 on the right).
The shock from the jet site is semi-relativistic (the velocity of plasma after the strong shock drops by a factor 
of three) and has relatively strong magnetic field. On the other hand, the shock in the GC wind is non-relativistic
with weak magnetic field. We expect that only shock in the jet plasma can accelerate efficiently particles to 
large energies transferring significant part of the jet power to relativistic electrons.
The power which can be extracted by the shock from the jet plasma can be estimated on, 
\begin{eqnarray}
L_{\rm sh}^{\rm GC} = L_{\rm j}({{R_{\rm sh}^{\rm GC}}\over{\theta l}})^2\approx 8.6\times 10^{38}(L_{36}M_{-5})^{1/2}~~~{\rm {{erg}\over{s}}}. 
\label{eq9}
\end{eqnarray}
\noindent
Note that the power estimated above for the single collision of the GC with the jet is clearly larger than in the case of a single collision of the red giant star with the jet but it becomes of the order of the power expected in the collision of a single massive star with the jet. Therefore, collisions of GCs with jets at the distance scale of a few kpcs from the central engine can be also responsible for the formation of powerful shocks in jets at such large distances. In the next section we discuss possible acceleration of electrons on the shocks formed in collisions of compact objects with jet plasma in a relatively nearby active galaxies.

\section{Acceleration of electrons at the shock in the jet}

In order to estimate the parameters of particles which could be eventually accelerated in such scenario, we consider
the physical conditions within the jet.
The upper limit on the magnetic field at the base of the jet can be estimated by assuming that the Poynting flux
($L_{\rm P}$) through the jet, is a part ($\mu$) of the total jet power ($L_{\rm j}$), i.e. 
\begin{eqnarray}
L_{\rm P} = \pi r_{\rm in}^2U_{\rm B}c\Gamma^2 = \mu L_{\rm j} = 10^{43}\mu L_{43}~~~{\rm erg~s^{-1}}, 
\label{eq10}
\end{eqnarray}
\noindent
where the inner radius of the jet can be related to the SMBH mass $r_{\rm in} = 3r_{\rm Sch} = 10^{14}M_8$ cm, the mass of the BH is $M_{\rm BH} = 10^8M_8$, the energy density of the magnetic field is $U_{\rm B} = B^2/8\pi$, and 
$\Gamma$ is the Lorentz factor of the jet. The jet can be either Poynting flux dominated (then $\mu\sim 1$) or 
matter dominated (then $\mu \ll 1$). The magnetic field at the base of the jet can be estimated by reversing Eq.~(\ref{eq10}), 
\begin{eqnarray}
B_{\rm b}\approx 520 (\mu L_{43})^{1/2}/(\Gamma M_8)~~~{\rm Gs}.
\label{eq11}
\end{eqnarray}
\noindent
The perpendicular component of the magnetic field in the jet (in respect to the jet axis) at the distance, $l$, from its base can be approximated by,
\begin{eqnarray}
B(l)\approx {{B_{\rm b} r_{\rm in}}\over{(r_{\rm in} + \theta l)}}\approx 
1.7\times 10^{-4}{{(\mu L_{43})^{1/2}}\over{\Gamma \theta_{-1}l_3}}~~~{\rm Gs}, 
\label{eq12}
\end{eqnarray}
\noindent
for large distances from the base of the jet. In this case, $B(l)$ does not depend on the mass of the black hole. 

We assume that electrons are accelerated at the shock from the site of the jet plasma to energies limited by
their synchrotron energy losses. The energy gain rate by electrons is assumed to be determined by the so called acceleration coefficient, $\xi = 10^{-3}\xi_{-3}$, 
\begin{eqnarray}
\dot{E}_{\rm acc} = \xi cE/R_{\rm L}\approx 9.1\times 10^{9}\xi_{-3} B(l)~~~{\rm eV~s^{-1}}, 
\label{eq13}
\end{eqnarray}
\noindent
where $E = 1E_{\rm TeV}$ TeV is the energy of electrons and $R_{\rm L}$ is the Larmor radius of electrons.
This acceleration process can be saturated by the energy losses of electrons.
The synchrotron energy loss rate is, 
\begin{eqnarray}
\dot{E}_{\rm syn} = {{4}\over{3}}c\sigma_{\rm T}U_{\rm B}(l)({{E}\over{m_{\rm e}c^2}})^2
\approx 75{{\mu L_{43}E_{\rm TeV}^2}\over{(\Gamma \theta_{-1}l_3)^2}}~~~{\rm {{eV}\over{s^{1}}}}, 
\label{eq14}
\end{eqnarray}
\noindent
where the energy density of magnetic field is, 
\begin{eqnarray}
U_{\rm B}(l) = B^2/(8\pi)\approx 740\mu L_{43}/(\Gamma \theta_{-1}l_3)^2~~~{\rm  {{eV}\over{cm^{3}}}},
\label{eq15}
\end{eqnarray}
\noindent
$\sigma_{\rm T}$ is the Thomson cross section and $m_{\rm e}$ is the electron rest mass. 
Then, the maximum energy of electrons, limited by the synchrotron energy losses, are
\begin{eqnarray}
E_{\rm syn}\approx 55({{\xi}\over{B}})^{1/2}\approx 
135{{(\xi_{-3}\theta_{-1}l_3\Gamma )^{1/2}}\over{(\mu L_{43})^{1/4}}}~~~{\rm TeV}. 
\label{eq16}
\end{eqnarray}
\noindent
Electrons with these maximum energies cannot have the Larmor radii larger than the dimension of the shock
around the compact objects. The conditions, $R_{\rm L} = R_{\rm sh}^{\rm GC}$ and $R_{\rm L} = R_{\rm sh}^\star$, introduce another absolute limit on the maximum electron energy, equal to $E_{\rm L}^{\rm GC}\approx 1.4\times 10^5\mu^{1/2}(L_{36}M_{-5})^{1/4}/\Gamma$ TeV and 
$E_{\rm L}^\star\approx 1.8\times 10^3(\mu M_{-7}v_3)^{1/2}/\Gamma$ TeV. The limits due to the Larmor radius are clearly above the maximum energies of accelerated electrons. However, another limit on the maximum energies of electrons is introduced by the time scale spend by electrons at the shock region. This limit differ for stars and GC.
Therefore, we consider it separately.

\subsection{Stars in jet}

The maximum energies of electrons can be also limited by their advection time scale along the shock with the flowing plasma. This time scale can be estimated from, 
\begin{eqnarray}
\tau_{\rm adv} = {{3R_{\rm sh}^\star}\over{\beta c}}\approx  3.6\times 10^6{{(M_{-7}v_3)^{1/2}\theta_{-1}l_3}\over{\beta L_{43}^{1/2}}}~~~{\rm s}, 
\label{eq17}
\end{eqnarray}
\noindent
where $\beta$ is the velocity of the jet plasma in units of the velocity of light. The comparison of the acceleration time scale ($\tau_{\rm acc} = E/\dot{E}_{\rm acc}$, see Eq.~\ref{eq13}) with the advection time scale, $\tau_{\rm adv}$, gives us the limit on the maximum energies of electrons accelerated in the stellar shock,
\begin{eqnarray}
E_{\rm adv}^\star\approx 5.6\xi_{-3}(\mu M_{-7}v_3)^{1/2}/(\beta\Gamma)~~~{\rm TeV}.
\label{eq18}
\end{eqnarray}
\noindent
Note that advection limit on the energy of electrons mainly depends on the parameters of the stellar wind. On the other hand, the synchrotron limit mainly depends on the parameters of the jet. In general, the advection limit becomes important in respect to the synchrotron limit for the red giant stars in contrast to the case of the massive stars.

Electrons accelerated at the shock around the star can also lose efficiently energy on the interaction with the stellar radiation. In order to evaluate importance of this process we calculate 
the energy density of stellar photons at the shock region, $U_\star = \sigma_{\rm SB}T_\star^4
(R_\star/R_{\rm sh}^\star)^2\approx 1.7\times 10^4T_3^4R_{12}^2L_{43}/[M_{-7}v_3(\theta_{-1}l_3)^2]$ eV cm$^{-3}$,
where the surface temperature of the red giant type star is $T_\star = 3.7\times 10^3T_3$ K and its radius 
$R_\star = 5\times 10^{12}R_{12}$ cm, and $\sigma_{\rm SB}$ is the Stefan-Boltzmann constant. On the other hand, energy density of the magnetic field at the shock from the jet site is given by Eq.~(15). The relative energy losses of electrons on these two processes (in the Thomson regime, i.e. for $E_{\rm e} \ll 280/T_3$ GeV) depends only on the ratio of the energy densities of the radiation and the magnetic field. Therefore, electrons with energies clearly below $\sim 300$ GeV will lose energy more efficiently on the IC process than on the
synchrotron process. However, the energy losses of electrons becomes comparable for TeV energies to which these particles are expected to be accelerated at the shocks provided that the jet is Poynting flux dominated ($\mu\sim 1$).
Since the winds of massive stars are clearly more extreme than those of red giants, a relatively stronger IC $\gamma$-ray emission (in respect to the synchrotron emission) is expected in the case of massive stars within the jets.

The power in relativistic electrons extracted from the shock, created in collision of the jet  with stellar winds, is estimated on,
\begin{eqnarray}
L_{\rm e}^\star = \eta N_\star L_{\rm sh}^\star\approx  1.45\times 10^{40}\eta_{-1}N_{6} M_{-7}v_3~~~{\rm {{erg}\over{s}}},
\label{eq19}
\end{eqnarray}
\noindent
where $\eta = 0.1\eta_{-1}$ is the energy conversion efficiency from the shock to relativistic electrons, and $N_\star = 10^6N_6$ is the number of stars within the jet. 
The energy conversion efficiency, $\eta$, is usually expected to be in the range $\sim$0.1-0.2.
The number of massive stars within the jet is expected to be a few orders of magnitudes lower than the number of red giants (see Sect.~2). However the product of the mass loss rate and the wind velocity of a massive star is typically 3-4 orders of magnitudes larger (see Sect.~3.1). As a result, the power in relativistic electrons accelerated on the shocks formed by the whole population of red giants and massive stars is expected to be on the similar level.

\subsection{Globular clusters in jet}

As above, we constrain the maximum energies of electrons due to their advection along the shock around the GC within the jet.
We compare the acceleration time scale,  $\tau_{\rm acc}$, with $\tau_{\rm adv} = 3R_{\rm sh}^{\rm GC}/\beta c$. This condition limits energies of electrons to,  
\begin{eqnarray}
E_{\rm adv}^{\rm GC}\approx 430\xi_{-3}\mu^{1/2}(L_{36}M_{-5})^{1/4}/(\beta\Gamma)~~~{\rm TeV}.
\label{eq20}
\end{eqnarray}
\noindent
In general, this limit is less restrictive than the advection limit for stars within the jet, since the shock dimensions
around GCs are expected to be clearly larger for typical parameters of the scenario.
The synchrotron limit on the acceleration of electrons dominates over the advection limit for the following condition on the least constrained parameter of the model, i.e.
the acceleration coefficient should be, $\xi_{-3} > 0.1\theta_{-1}l_3\beta_1^2\Gamma^3/(L_{43}L_{36}M_{-5})^{1/2}/\mu^{3/2}$.
If the above condition is fulfilled, then we can  estimate the characteristic energies of synchrotron photons produced by electrons on $\varepsilon_{\rm x}\approx m_{\rm e}c^2 (B/B_{\rm cr})(E_{\rm syn}/m_{\rm e}c^2)^2 \approx
1.4\times 10^5\xi$ keV, where $B_{\rm cr} = 4.4\times 10^{13}$ Gs is the critical magnetic field strength. In the case of Cen A, the X-ray emission up to $\sim 10$ keV is observed (Hardcastle et al.~2006). Therefore, the acceleration coefficient is expected to be at least $\xi\geq 10^{-4}$. Electrons with such energies should produce synchrotron emission extending to X-ray energies but also TeV $\gamma$-rays by scattering optical radiation produced within the galactic bulge and the MBR.

Let us check whether electrons, accelerated at the shock around GC, will manage to cool efficiently on the synchrotron or the Inverse Compton (IC) processes. The electron cooling mean free path on the synchrotron process in the magnetic field of the jet is, 
\begin{eqnarray}
\lambda_{\rm syn} = {{Ec}\over{\dot{E}_{\rm syn}}}\approx {{0.13(\Gamma \theta_{-1}l_3)^2}\over{(\mu L_{43}E_{\rm TeV})}}~~~{\rm kpc}. 
\label{eq21}
\end{eqnarray}
\noindent
The synchrotron cooling process of electrons will occur locally in  the jet, provided that the synchrotron mean free path becomes shorter than the characteristic distance scale which we identify with the distance of the injection place of electrons from the base of the jet, i.e. $\lambda_{\rm syn} < l$.
The above condition is fulfilled for electrons with  energies $E > 0.1\theta_{-1}^2l_3/(L_{43}\Gamma)$ TeV. 
The IC energy losses of electrons can in principle be determined by the soft radiation produced by the GC itself, the galactic bulge thermal radiation and the MBR. The bulge radiation might play an important role in the inner part of the intermediate scale jet (i.e $\sim$kpc scale). The MBR can start to be important
in the outer parts of the large scale jet when the energy density of the magnetic field drops significantly. 
The energy density of radiation  from the GC at the location of the shock can be estimated on
$U_{\rm GC} = L_{\rm GC}/4\pi c(R_{\rm sh}^{\rm GC})^2\approx 320L_{6}L_{43}/(L_{36}M_{-5})^{1/2}(\theta_{-1}l_3)^2$ eV 
cm$^{-3}$, where the optical luminosity of the GC is $L_{\rm GC} = 10^6L_6L_\odot = 4\times 10^{39}$ 
erg s$^{-1}$ and $L_\odot$ is the luminosity of the Sun.
The energy density of the bulge radiation is estimated on $U_{\rm bulge} = 
L_{\rm bulge}/(4\pi R_{\rm bulge}^2c)\approx 70L_{11}/R_3^2$ eV cm$^{-3}$, where the bulge luminosity is $L_{\rm bulge} = 10^{11}L_{11}$ L$_\odot$, and the bulge radius is $R_{\rm bulge} = 1R_3$ kpc. The energy density of the MBR is $U_{\rm MBR} = 0.25$ eV cm$^{-3}$. The mean free path for electron energy losses on IC process in the Thomson regime in these radiation fields can be estimated on,
\begin{eqnarray}
\lambda_{\rm IC}^{\rm T} = cE/\dot{E}_{\rm IC}\approx 94/(E_{\rm TeV}U_{\rm eV})~~~{\rm kpc}, 
\label{eq22}
\end{eqnarray}
\noindent
where $U_{\rm rad} = 1U_{\rm eV}$ eV cm$^{-3}$. 
We expect that multi-TeV electrons will mainly scatter stellar radiation in the Klein-Nishina regime but the estimate based on the Thomson cross section allows us to have impression about importance of the IC energy losses.
The IC losses becomes important when the mean free path of electrons is comparable to the characteristic distance scale on which they propagate. In the case of the radiation field from the GC, $R_{\rm sh}^{\rm GC}$ is always clearly lower than  $\lambda_{\rm IC}^{\rm T}$, for reasonable parameters of the scenario. 
Therefore, energy losses of electrons in the GC radiation can be safely neglected. In the case of the bulge radiation, the mean free path of the IC energy losses can become comparable to the characteristic size of the bulge (of the order of kpc) and also to the mean free path of electrons on the synchrotron process. Therefore, the IC energy losses in the bulge radiation should be taken into account on the kpc distance scale from the base of the jet. 
The mean free path for the IC energy losses of electrons in the MBR is of the order of
$\sim$kpcs, for electrons with energies of a hundreds TeV. Therefore, IC losses of electrons are expected to become important in the outer parts of the jets at the distance scale of several kpc from the base of the jet. Note that the GC halo still extends to such distances.

We conclude that in fact, electrons can be accelerated to large energies at the shocks formed in collisions 
of GCs with jet plasma. They should produce synchrotron radiation extending up to X-ray energy range and also multi-TeV
$\gamma$-rays by scattering radiation from the galactic bulge and the MBR.

The total power in relativistic electrons, accelerated at the shocks around GCs, can be estimated on, 
\begin{eqnarray}
L_{\rm e}^{\rm GC} = \eta N_{\rm GC} L_{\rm sh}^{\rm GC} \approx 8.6\times 10^{38}\eta_{-1}N_{1} 
(L_{36}M_{-5})^{1/2}~~~{\rm {{{erg}}\over{s}}}, 
\label{eq23}
\end{eqnarray}
\noindent
where the shock power from a single GC is given by Eq.~\ref{eq9},  and $N_{\rm GC} = 10N_1$ is the number of GCs within the jet.

The maximum energies of electrons are limited in this case by their synchrotron energy losses to values given by 
Eq.~(16).
For the expected values of the parameters describing considered scenario, the power in electrons, accelerated at the shocks around GCs, is estimated to be about an order of magnitude lower than in the case of the shocks around stars. However, electrons, accelerated at the shocks around GCs, are injected into the jet at much larger distances from its base (i.e. above $\sim$kpc). At these distances the jet magnetic field strength is already weak.  Therefore,  electrons lose relatively more efficiently 
energy on the IC process (producing $\gamma$-rays) than on the synchrotron process (producing radiation below X-ray energies). 

In the next section we perform detailed calculations of the multi-wavelength spectra expected from the intermediate scale jet assuming that different types of compact objects (red giants, massive stars, GCs) form multiple shocks in the jet plasma. As an example, the parameters of the nearby radio galaxy, Cen A, are applied.

\section{Non-thermal radiation from electrons}

We calculate expected synchrotron and IC spectra produced by electrons accelerated on the fronts of multiple
shocks within the jet for different parameters describing compact objects and the jet content. It is assumed that electrons are accelerated with  the differential power law spectra above some minimum energy $E_{\rm min}$. The spectral index equal to $-2$ is selected for the example calculations as expected in the shock acceleration scenario. The maximum  energies of electrons, 
$E_{\rm max}$, are determined either by the synchrotron energy losses (Eq.~\ref{eq16}) or by the escape from the acceleration region due to the advection of electrons with the jet plasma (Eq.~\ref{eq18} or~\ref{eq20}). 
These electron spectra have been normalized to the power transferred to electrons from the shocks in the jet (see Eqs.~\ref{eq19} and ~\ref{eq23}). In order to obtain the  synchrotron and the IC $\gamma$-ray spectra, we
inject electrons at different distances from the base of the jet and simulate their propagation in the jet magnetic field and the radiation fields, discussed above, by applying the Monte Carlo method. The Klein-Nishina effect has been taken into account when calculating the IC $\gamma$-ray spectra. In these example calculations, we assume that the jet is semi-relativistic, with the apparent velocity equal to $\beta_{\rm app} = 0.6$ and moves at relatively large angle to the observer's line of sight  estimated on $\alpha = 50^{\circ}$ in the case of Cen A.  For the above parameters the velocity of the jet is
estimated on $\beta\approx 0.52$ and its Lorentz factor on $\Gamma\approx 1.172$. The Doppler factor of the jet in Cen A at the distance of a hundred pc from its base is then estimated on  $D = 1/[\Gamma (1 -\beta\cos\alpha)]\approx 1.282$. 
Note that at the kpc distance scales the jet may significantly decelerate. Therefore, Doppler effects can be negligable in the example case of Cen A. Such assumption clearly simplifies the calculations of the non-thermal radiation presented in this paper.
The jet has a constant opening angle equal to $\theta = 0.1$ rad and the jet power is $L_{\rm j} = 10^{43}$ erg s$^{-1}$. For these basic parameters, we investigate the spectra as a function of two other parameters determining the relativistic electrons in the jet, the jet magnetization parameter $\mu$ and the acceleration efficiency of electrons in the jet $\xi$. These two parameters determine the relative part of energy of relativistic electrons lost on the synchrotron and on the IC process.

\begin{figure*}
\vskip 9.5truecm
\includegraphics{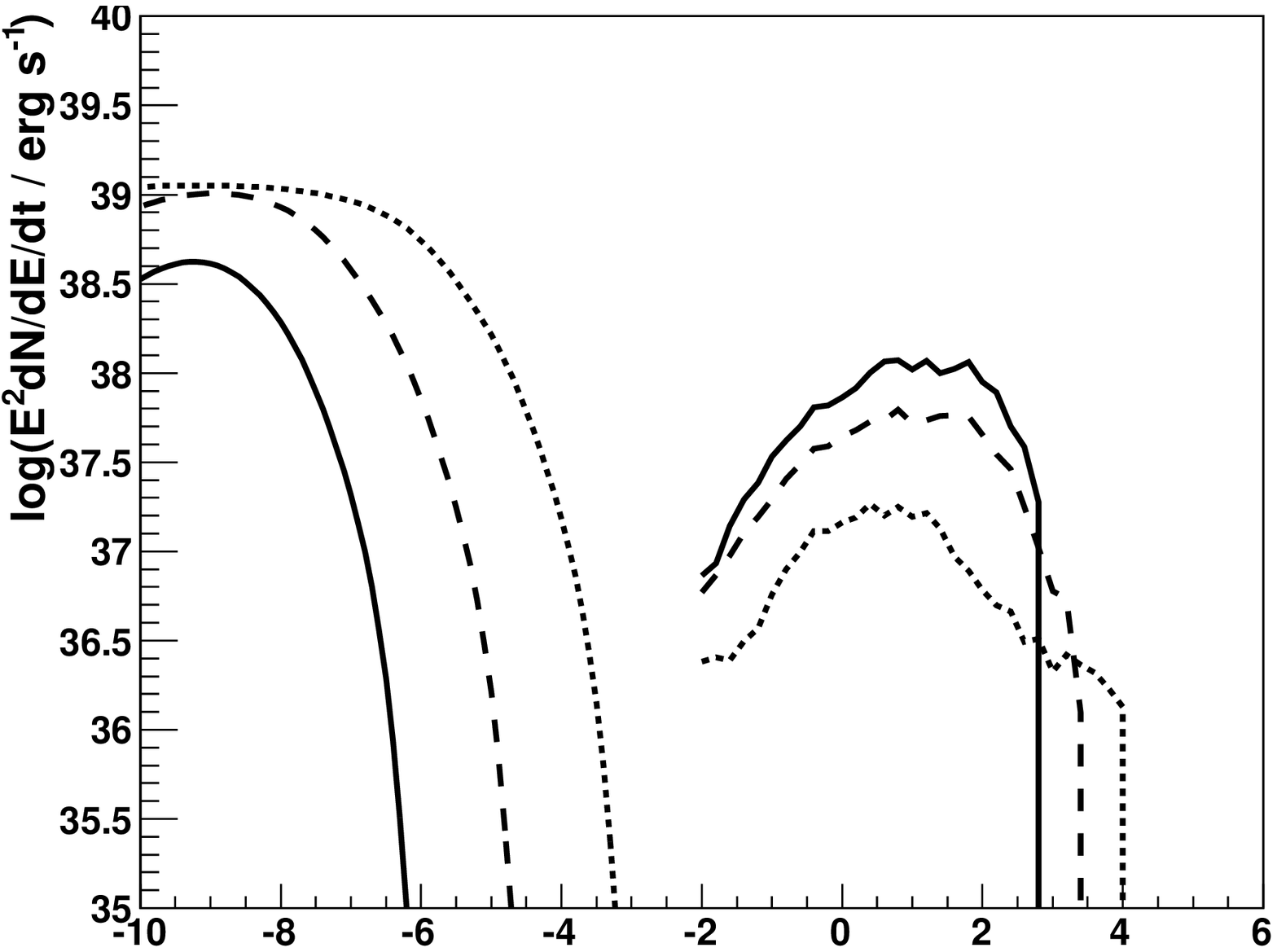}
\includegraphics{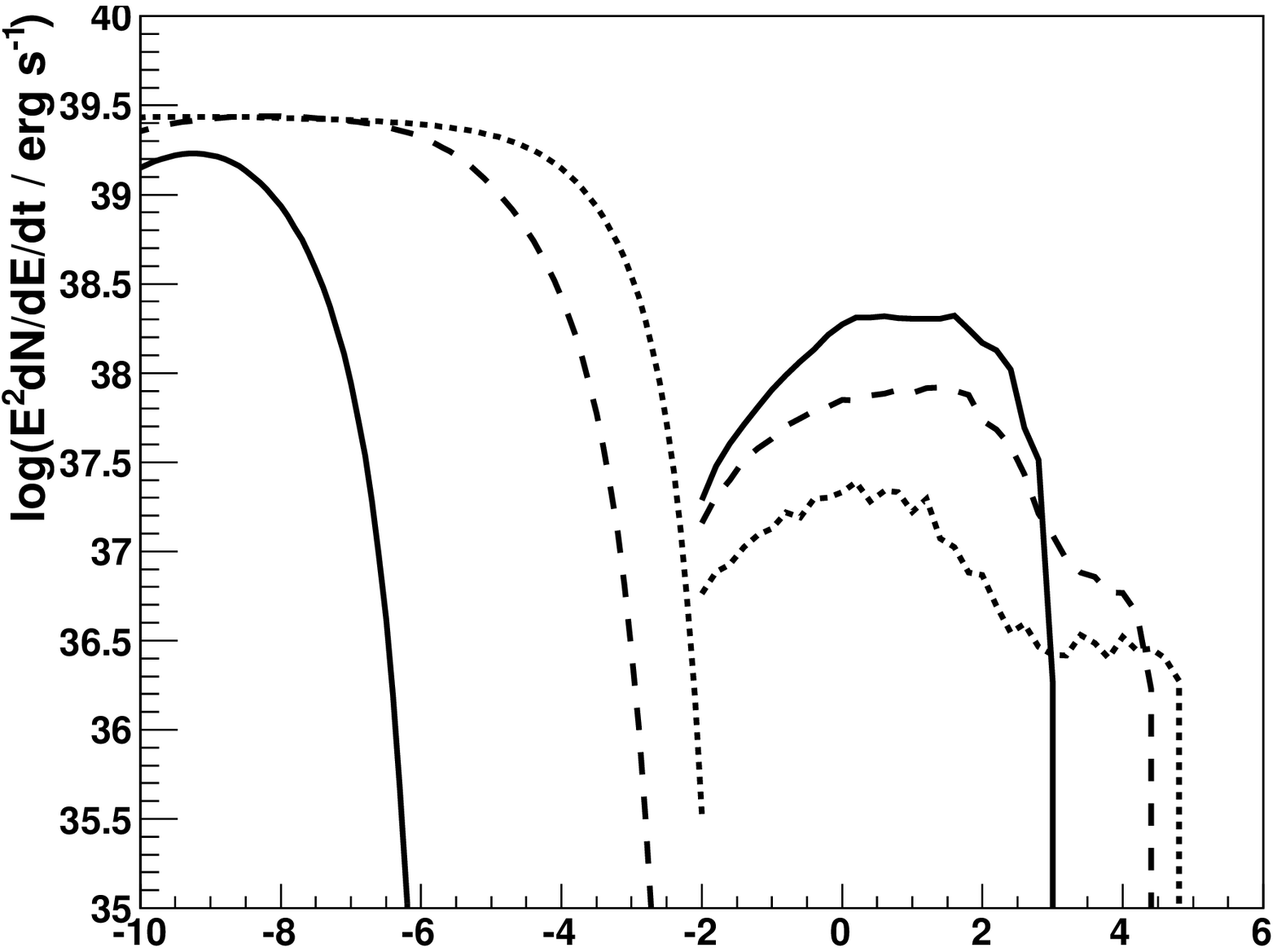}
\includegraphics{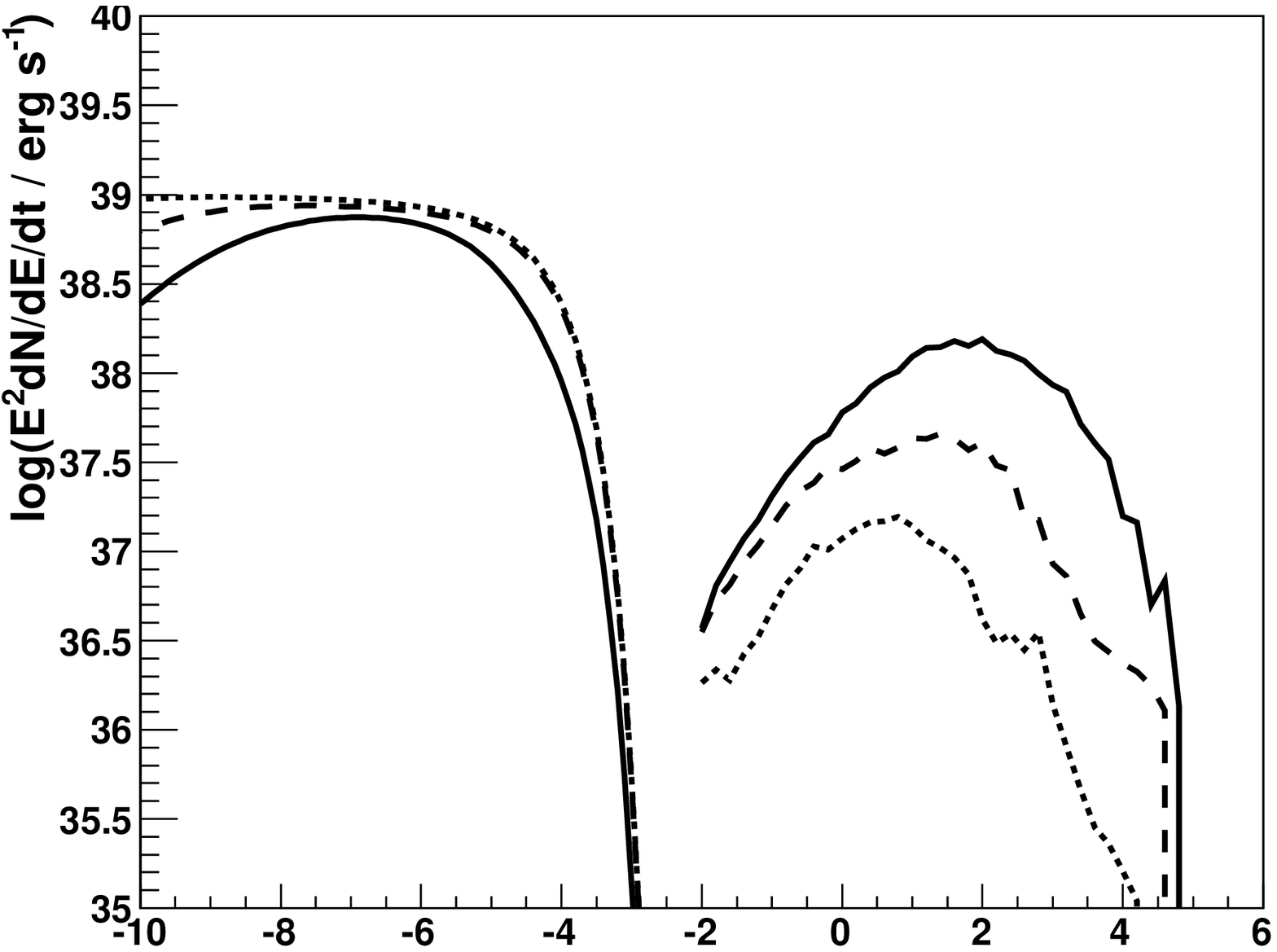}
\includegraphics{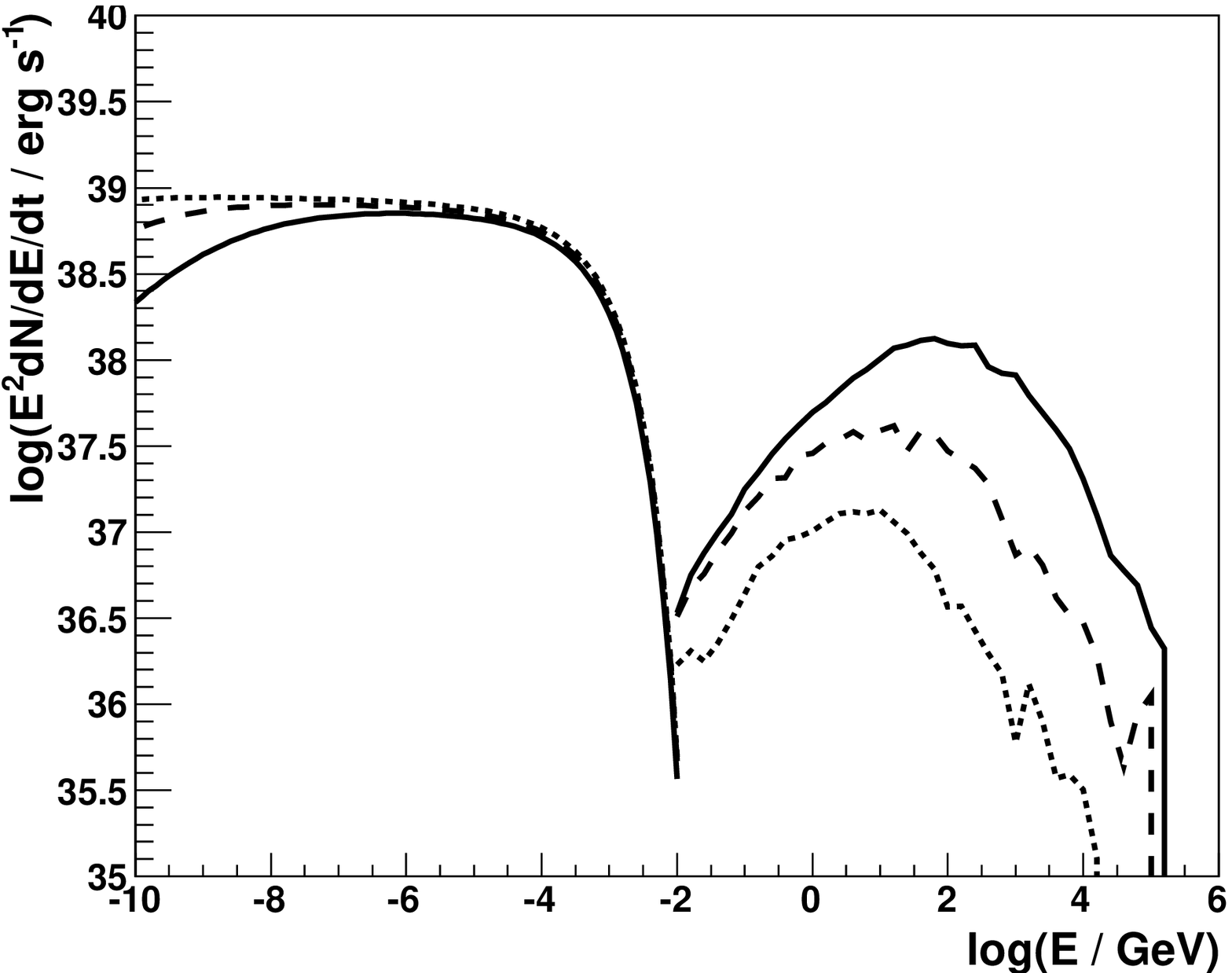}
\caption{The synchrotron and IC spectra (SED - spectral energy distribution) produced by electrons accelerated at the multiple shocks in the jet plasma around the red giants (on the left) and massive stars (on the right), which entered the jet homogeneously on the range of distances from the base of the jet, $L_{\rm min} = 10$ pc and $L_{\rm max} = 1$ kpc. The results are shown for the Poynting dominated jet, for the value of magnetization parameter equal to $\mu = 1$ (dotted curves)) and the matter dominated jets described by $\mu = 0.1$ (dashed), and $\mu = 0.01$ (solid). The upper panel shows the spectra for the acceleration parameter of electrons equal to $\xi = 10^{-3}$ and the bottom panel for $\xi = 10^{-2}$. The parameters of the jet are the following $L_{\rm j} = 10^{43}$ erg s$^{_1}$, $\theta = 0.1$ rad.  
It is assumed that $10^6$ red giants and 100 massive stars are present in the jet at the same time. The energy conversion efficiency from the jet to relativistic leptons is $10\%$. 
The other parameters of the stars and the radiation field are reported in the main text.}
\label{fig2}
\end{figure*}
\subsection{Emission from star-jet collisions}

The calculations of the IC $\gamma$-ray spectra in the case of collisions of stars with the inner jet (i.e. at distances $<1$ pc) have been already considered by e.g. Bednarek \& Protheroe~(1997). In such case, the shock is located relatively close to the massive companion star and relativistic electrons initiate cascade in the dense stellar
radiation field. The calculations of the synchrotron emission from electrons accelerated on the multiple shock structures around the stars in the intermediate scale jets has been recently considered by Wykes et al.~(2014). But no IC $\gamma$-ray emission has been calculated in this model. 

We consider the star-jet interaction process on a large distance scale from the base of the jet. In such case, the cascade process initiated by electrons accelerated around massive stars is inefficient since the shock is too far from the star. Electrons escape with significant energies from the shock and propagate mainly in the jet volume. At this stage, electrons lose energy on the synchrotron process in the magnetic field of the jet and on the IC process by scattering different radiation fields such as radiation of compact star, diluted radiation from the galactic bulge and the MBR. Electrons propagating outside the bulge, with typical dimension of $R_{\rm bulge}$ = 1 kpc, will see the radiation field diluted by the factor $(l/R_{\rm bulge})^2$.  
In these example calculations we assume that electrons are accelerated on shocks around stars entering uniformly the jet 
between 10 pc up to 1 kpc. Note that in such case maximum energies of electrons injected from the shocks into the jet depend on the distance from the base of the jet as given by Eq.~\ref{eq16} or Eq.~\ref{eq18}. The parameters of stars have been fixed on $\dot{M}_{\rm w} = 10^{-7}$ M$_\odot$ yr$^{-1}$ and $v_\star = 30$ km s$^{-1}$ (for the red giant stars) and on $\dot{M}_{\rm w} = 10^{-5}$ M$_\odot$ yr$^{-1}$ and $v_\star = 10^3$ km s$^{-1}$ (for the massive stars).

Results of these example calculations of the synchrotron and IC $\gamma$-ray spectra are shown in Fig.~2. We investigate the spectra for the Poynting flux dominated jets (the magnetization parameter of the jet $\mu = 1$) and for the matter dominated jets (described by $\mu = 0.1$ and $0.01$). We also show the spectra obtained for different values of the acceleration efficiency of electrons $\xi = 10^{-2}$ and $10^{-3}$. In all considered cases, the synchrotron spectra dominate over the IC spectra. As expected, the IC $\gamma$-ray spectra are stronger for less magnetized jets. In the case of red giants, the cut-off in synchrotron spectrum is determined by the advection time scale of electrons along the shock structure. In contrast, the synchrotron spectra cuts-off independently on the jet magnetization for shocks around massive stars.

\subsection{Emission from globular cluster-jet collisions}

Collisions of GCs with jets in active galaxies have not been considered up to now as a source of disturbances in the jet plasma (shocks) which might be responsible for acceleration of particles. 
However, a large number of GCs around active galaxies, and evidences that their numbers are correlated with the 
mass of the central black hole (Harris et al.~2014), suggest that significant number of GCs is immersed in the jet.
We showed above that shocks around GCs provide conditions for acceleration of electrons to hundreds of TeV.
Applying the numerical code for the propagation of electrons in the jet and their radiation described above, we calculate the synchrotron and IC spectra. Since such calculations have not been considered before, we also show 
the results of calculations of the spectra produced at different distances from the base of the jet. 
The spectra are investigated as a function of parameters describing the acceleration process of electrons (i.e.
the magnetization parameter of the jet $\mu$ and the acceleration efficiency of electrons $\xi$). Electrons
are injected at fixed distance from the base of the jet and, after leaving the shock, they are frozen in the jet plasma. In such case, the synchrotron energy losses of electron are determined by the profile of the magnetic field along the jet (see Eq.~\ref{eq12}). On the other hand, these electrons also comptonize radiation from the galactic bulge (with photon density dropping with the distance along the jet) and the MBR. Note that electrons are accelerated to larger energies farther along the jet.
As in the case of collisions of stars with the jet, we assume that electrons reach the power law spectrum extending up to the maximum energies described above. In the case of GCs, the minimum energy of electrons is fixed on  1 TeV which corresponds to the characteristic energy of leptons injected by the millisecond pulsars into the mixed pulsar/stellar wind from the GC. 

\begin{figure*}
\vskip 8.5truecm
\includegraphics{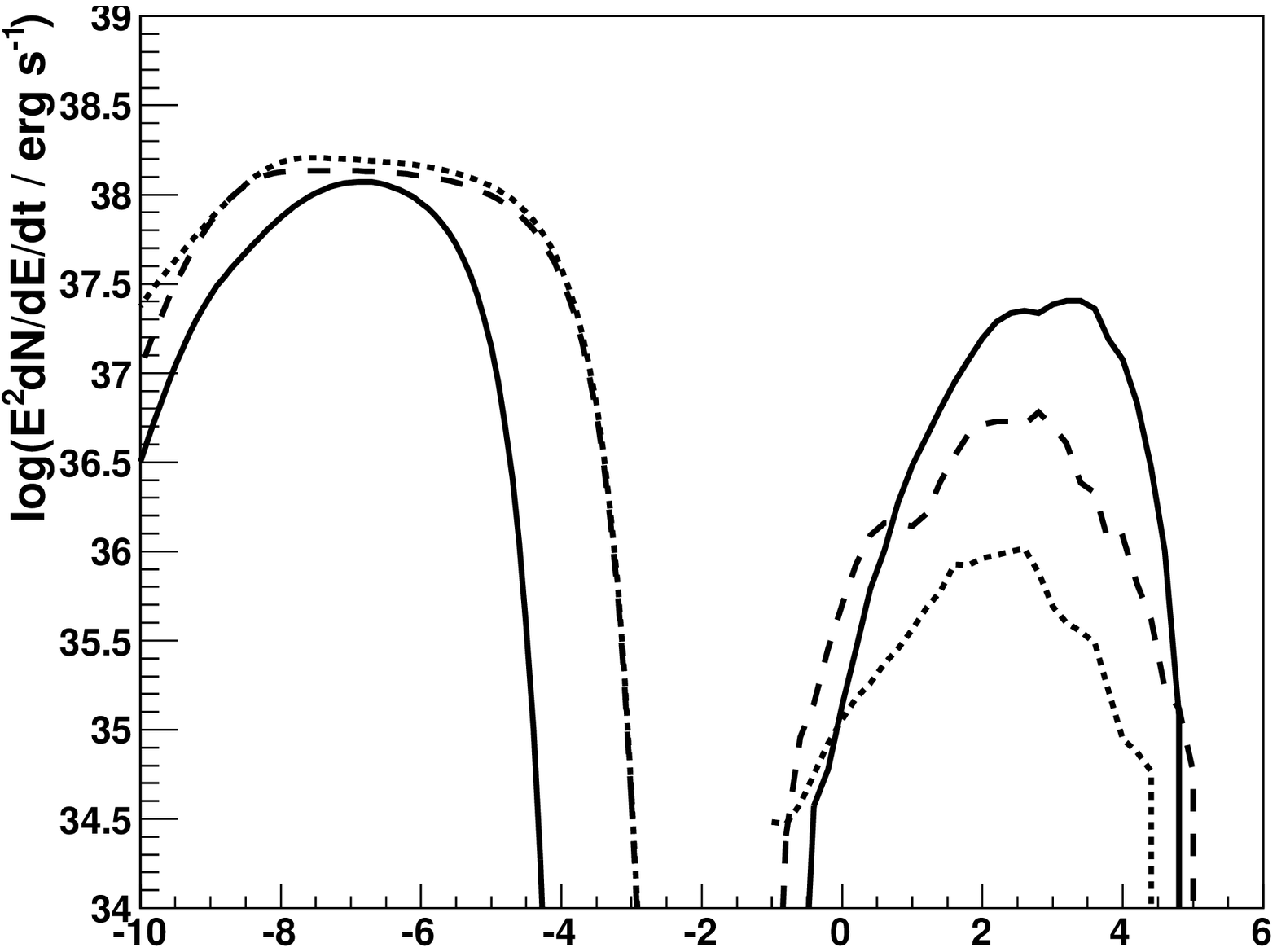}
\includegraphics{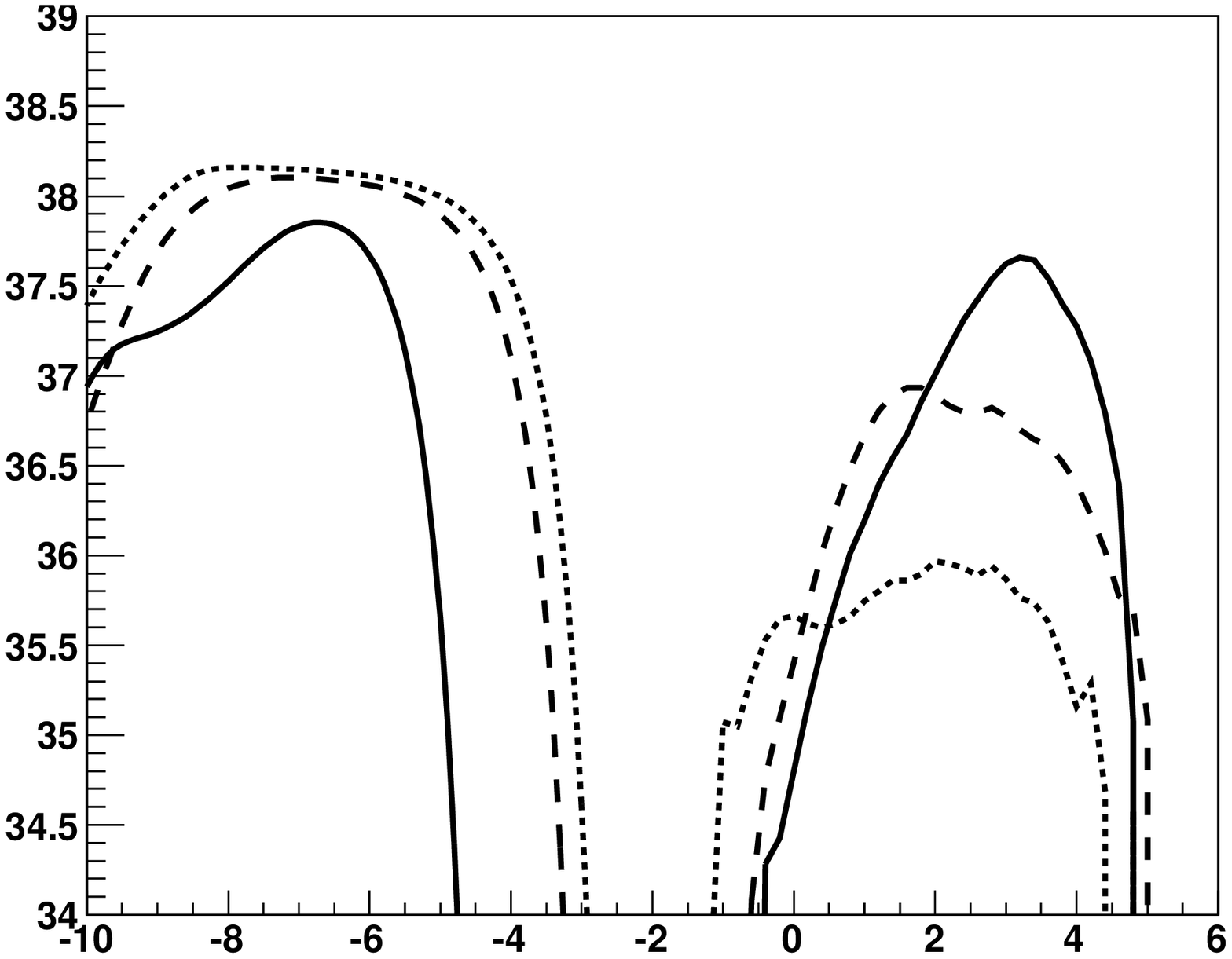}
\includegraphics{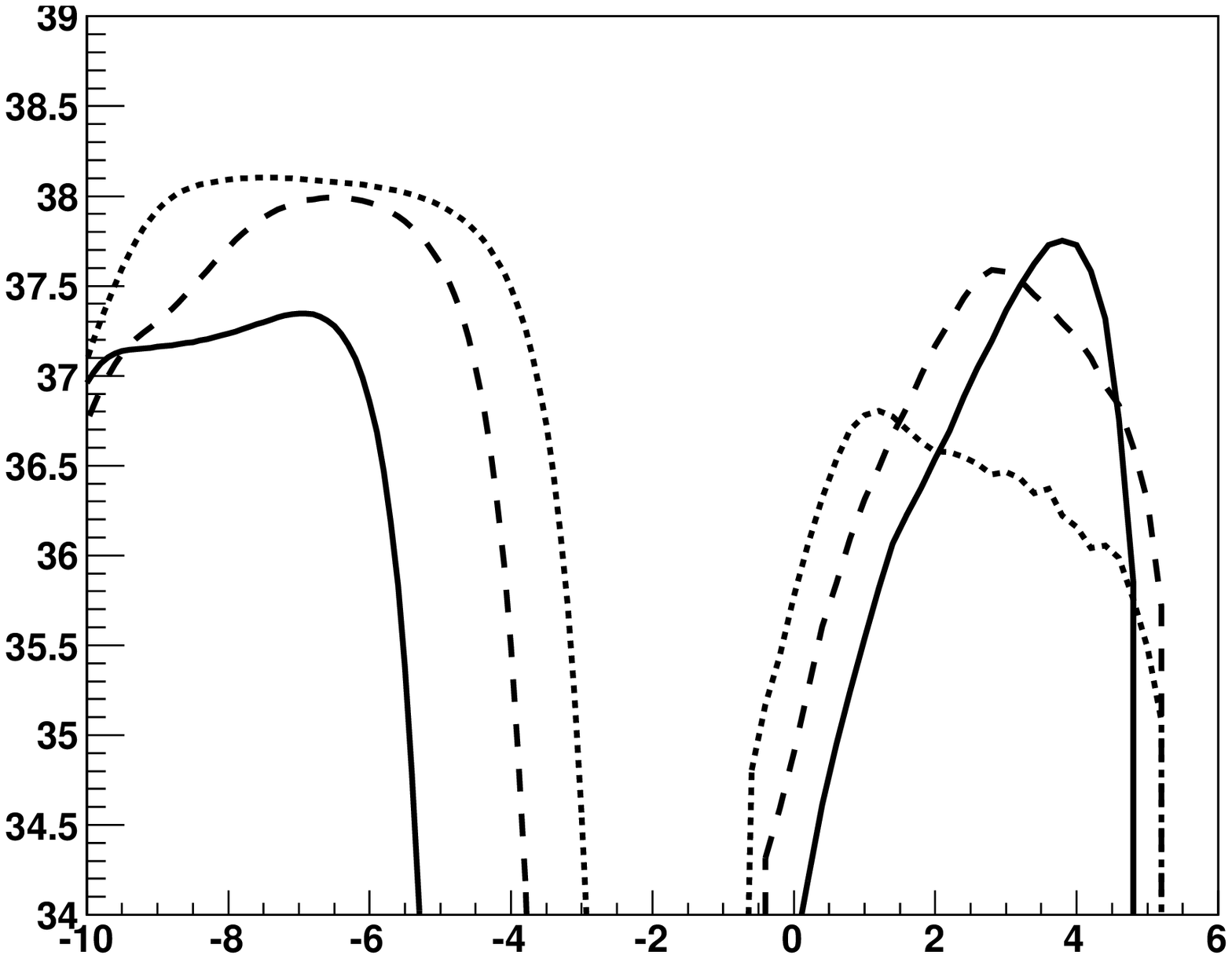}
\includegraphics{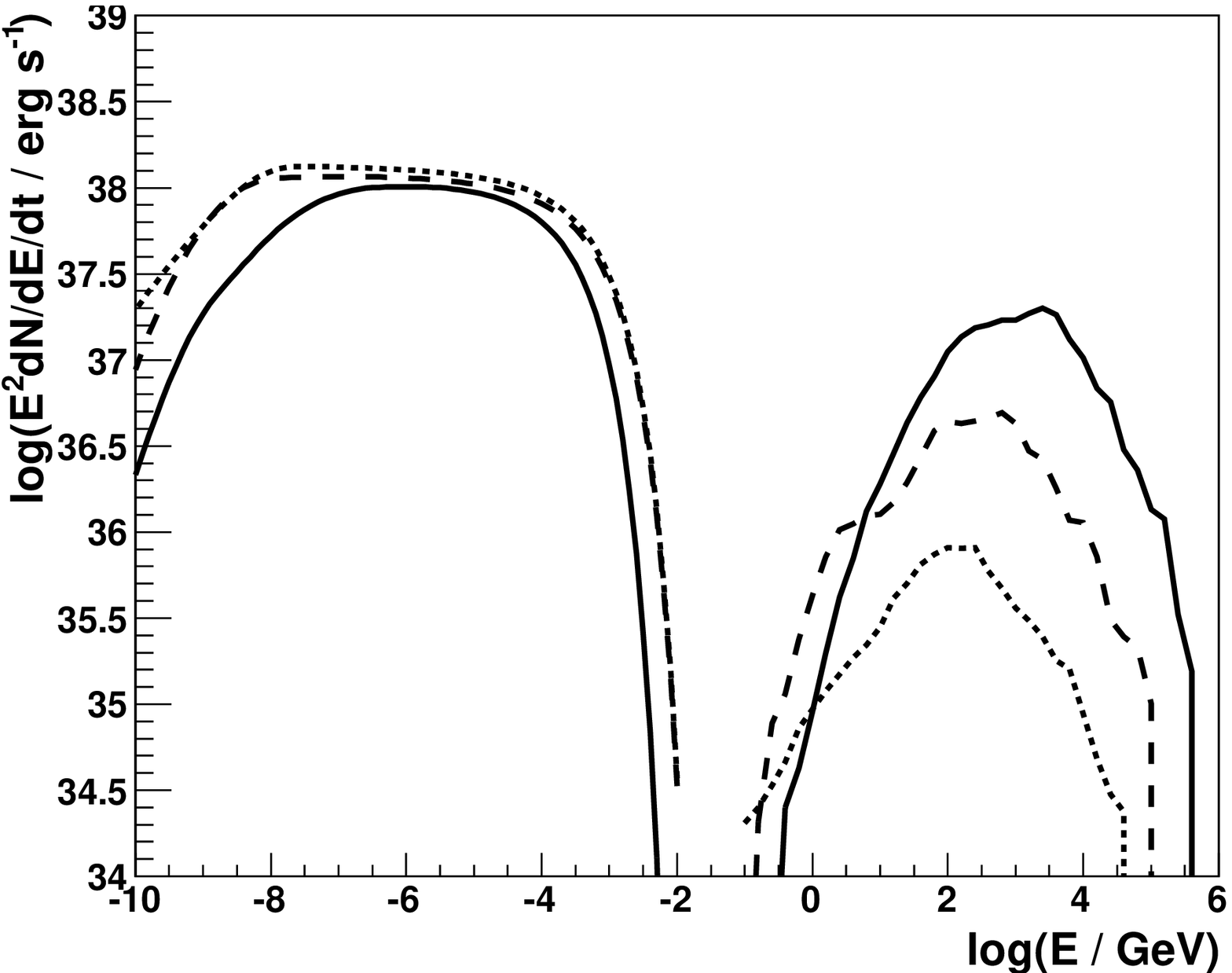}
\includegraphics{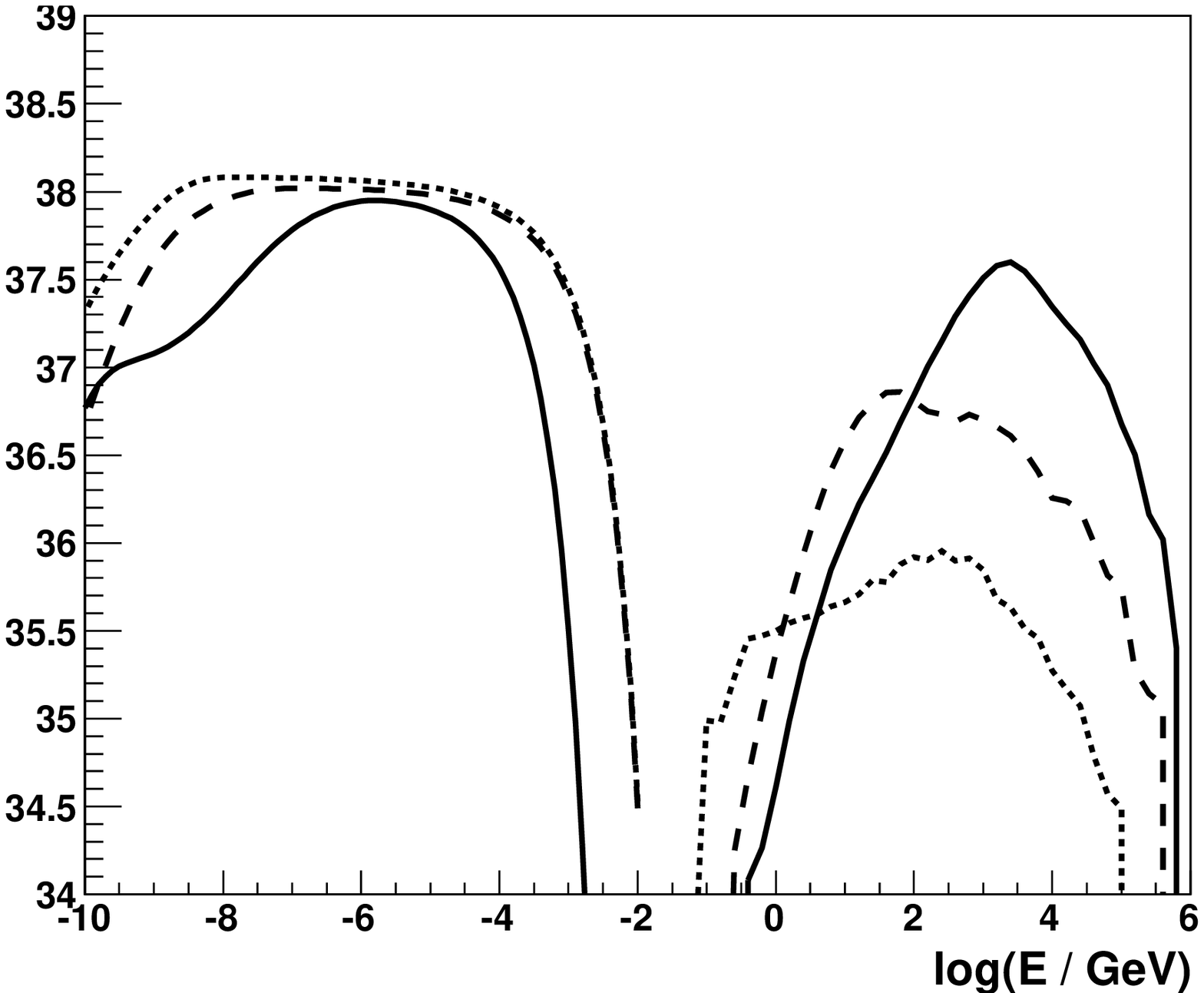}
\includegraphics{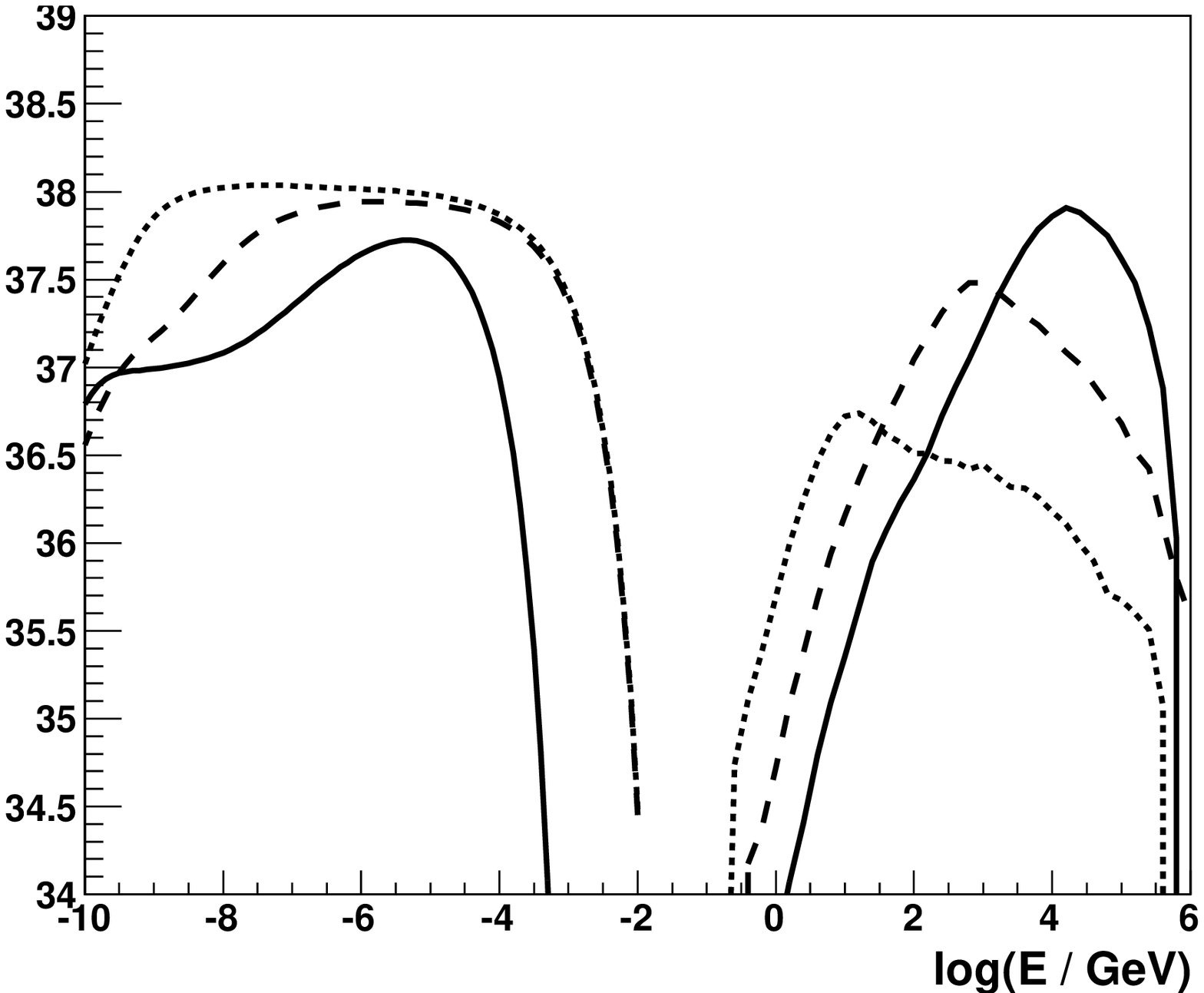}
\caption{The synchrotron and IC spectra (SED) produced by leptons accelerated at the shock in the jet plasma around the GC which enters the jet at different distances from its base, $l = 1$ kpc (on the left), 3 kpc (middle), and 10 kpc (on the right). The results are shown for the Pointing dominated jet,
for the value of magnetization parameter equal to $\mu = 1$ (dotted curves), and the matter dominated jets for $\mu = 0.1$ (dashed), and $\mu = 0.01$ (solid). The upper panel is for the acceleration parameter of electrons equal to $\xi = 10^{-3}$ and the bottom panel for $\xi = 10^{-2}$. The parameters of the jet are as in Fig.~2. It is assumed that 20 GCs are present in the jet and the energy conversion efficiency from the jet to relativistic leptons is $10\%$. The parameters describing the GC and the radiation field of the bulge are reported in the main text.}
\label{fig3}
\end{figure*}

In the example calculations of the synchrotron and the Inverse Compton spectra, we apply the parameters
derived for the nearby radio galaxy Cen A: the mass of the black hole $M_8 = 0.5$, the jet power $L_{43} = 1$, the jet opening angle $\theta_{-1} = 1$, the jet velocity $\beta = 0.5$ ($\Gamma\approx 1.15\sim 1$), and the parameters of typical for GCs:
GC stellar luminosity $L_6 = 1$, the power supplied by MSPs $L_{36} = 3$, the mass loss rate of red giant stars
$M_{-5} = 3$, and the parameters of the host galaxy bulge, the bulge stellar luminosity $L_{\rm bulge} = 10^{11}L_\odot$, the bulge radius $R_3 = 1$. The acceleration coefficient is taken to be $\xi_{-3} = 1$ in order to be consistent with the observations of the non-thermal X-ray emission from the jet. The radius of the shock around GC is then $R_{\rm sh}\approx 2.7l_3$ pc, the magnetic field strength along the jet is 
$B(r)\approx 5.6\times 10^{-5}\mu^{1/2}/l_3$ Gs, the maximum energies of electrons due to synchrotron energy losses are $E_{\rm syn}\approx 170(l_3/\mu )^{1/2}$ TeV, the maximum energies of electrons due to advection from the shock are $E_{\rm adv}\approx 1800\mu^{1/2}$ TeV. The comparison of these two last limits gives us the range of distances for which synchrotron limit is more restrictive, i.e. $l_3 < 100\mu^2$. We conclude that the maximum electron energies are always limited by synchrotron losses in the case of Poynting dominated jet. They are limited by the advection process in the case of the matter dominated jet with $\mu = 0.01$. For $\mu = 0.1$ the acceleration process of electrons is limited by the synchrotron
energy losses in the inner part of the jet, i.e. below $l_3 = 1$, but in the outer parts of the jet by the 
advection process.
The mean free paths of electrons are $\lambda_{\rm syn}\approx 240l_3^2/(\mu E_{\rm TeV})$ pc on the synchrotron process, $\lambda_{\rm IC}^{\rm bulge}\approx 1.3/E_{\rm TeV}$ kpc on the IC scattering of bulge radiation , and 
$\lambda_{\rm IC}^{\rm MBR}\approx 380/E_{\rm TeV}$ kpc on the IC scattering of the MBR.
The characteristic distance scales for IC energy losses of electrons on the border between the Thomson (T) and the Klein-Nishina (KN) regimes in the bulge radiation are comparable to the bulge dimension. These distance scales for IC energy losses of electrons with the TeV energies in the MBR are comparable to the dimensions of jets in AGNs. 

In Fig.~3, we show the synchrotron and the Inverse Compton spectra for different distances from the base of the jet, 
$l = 1$ kpc, 3 kpc, and 10 kpc.
The Poynting flux dominated jets ($\mu = 1$) and the matter dominated jets ($\mu = 0.1$ and 0.01) are considered. In the case of the Poynting dominated jets, the IC spectra are on a much lower level than 
the synchrotron spectra especially at lower distances from the base of the jet. On the other hand, the IC spectra start to dominate over synchrotron spectra at larger distances from the base of the jet in the case of matter dominated jets
defined by the magnetization parameter $\mu\sim 0.01$. Therefore, most of the energy of relativistic electrons can be transferred to the TeV $\gamma$-rays in jets at distances of a few kpcs from its base. The synchrotron spectra, produced in the jet by these electrons, clearly extend up to the X-ray energy range, even for the advection dominated acceleration process of electrons, for the considered range of parameters, 
$0.01<\mu < 1$ and $10^{-3}<\xi < 10^{-2}$.  

The synchrotron and IC spectra, produced by electrons accelerated at shocks formed by GCs entering the jet at the range of distances from its base between 1 kpc and 20 kpc, are shown in Fig.~4. It is assumed that GCs enter the jet homogeneously over such range of distances and ejects electrons with the spectra and powers as described above. The results are shown for the parameters as in previous figures. It is clear that in the case of the matter dominated jets the level of the TeV $\gamma$-ray emission is comparable to the level of the synchrotron X-ray emission. In this case, the $\gamma$-ray spectra peak at multi-TeV energies and the synchrotron spectra extends above $\sim$10 keV.
However, in the case of Poynting dominated jets, the synchrotron emission clearly dominates over the TeV $\gamma$-ray emission. The $\gamma$-ray luminosity is expected to be about an order of magnitude below the X-ray luminosity and
the $\gamma$-ray spectra peak at GeV energies.

We show that in the case of matter dominated jet models, the collisions of GCs with the kpc scale jets can be responsible for the detectable synchrotron X-ray emission and also for the TeV $\gamma$-ray emission. The relative levels of the emission in these two energy ranges should provide important constrains on the content of jets at kpc scale distances. 
Note that in the matter dominated jets the maximum energies of accelerated electrons are determined by their escape from the shock region due to the advection process. These electrons are not completely cooled when propagating in the jet on 
the distance scale of 20 kpc considered in the paper. Therefore, the IC spectra are so hard below the peak at TeV energies.
The lower energy electrons are cooled on much longer distance scale. They should produce steeper spectra on similar angular scales in the case of distant AGNs for which the contribution to the IC spectrum comes from more extended part of the jet. Therefore, such hard spectra are only expected from the intermediate scale jets in the nearby AGNs such as Cen A for which the distance of 20 kpc corresponds to angular size of the source several arc min. We considered only semi-relativistic jets as observed from nearby AGNs. In the case of relativistic jets, expected in the distant AGNs, these maximum energies of accelerated electrons
are expected to be lower (see Eq.~20). Therefore, the spectra may not extend to $\sim$10 TeV. These spectra are not expected to be  
hard enough to provide interesting constraints on the Extra-galactic Background Light.

\begin{figure}
\vskip 9.5truecm
\includegraphics{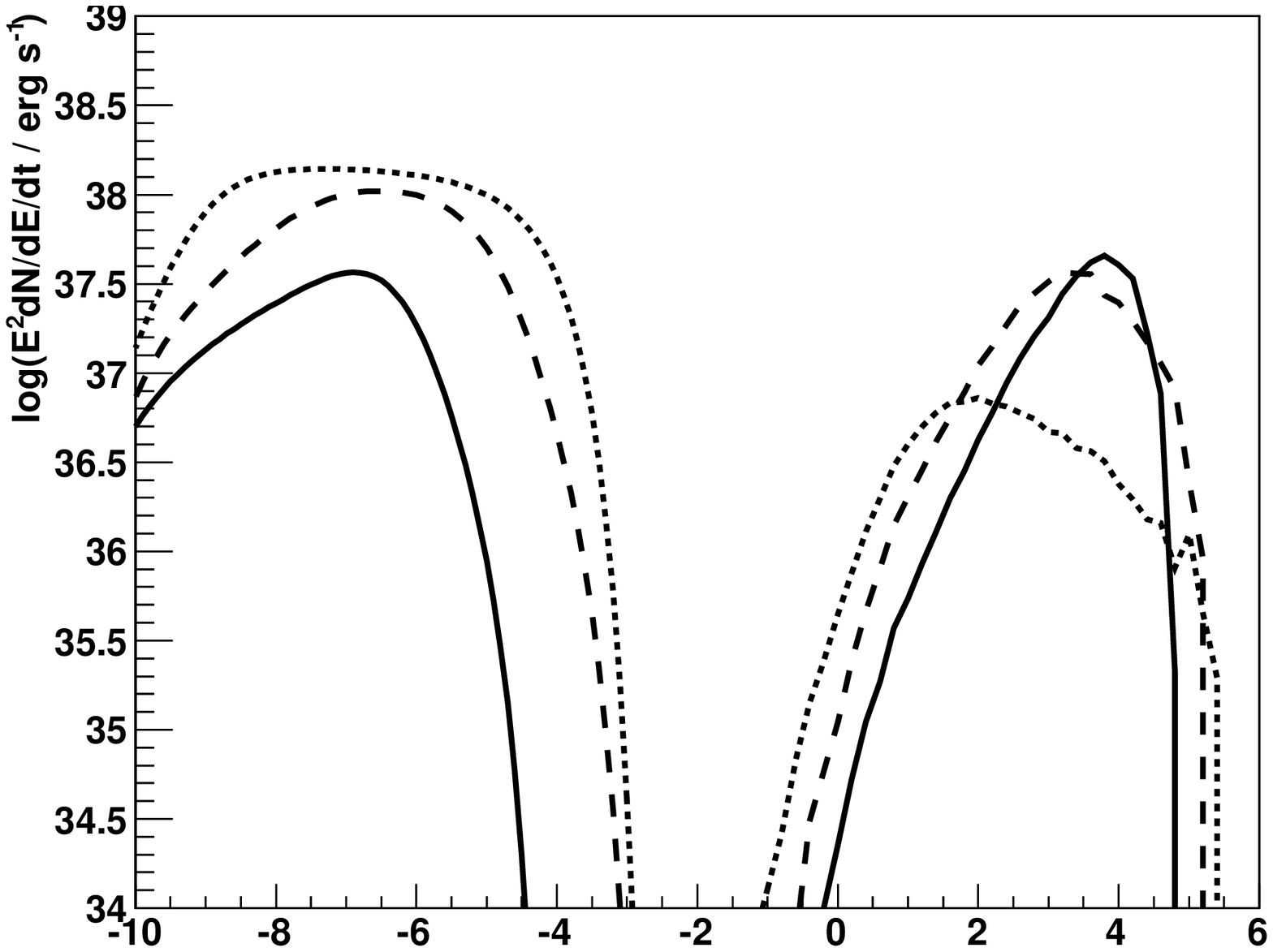}
\includegraphics{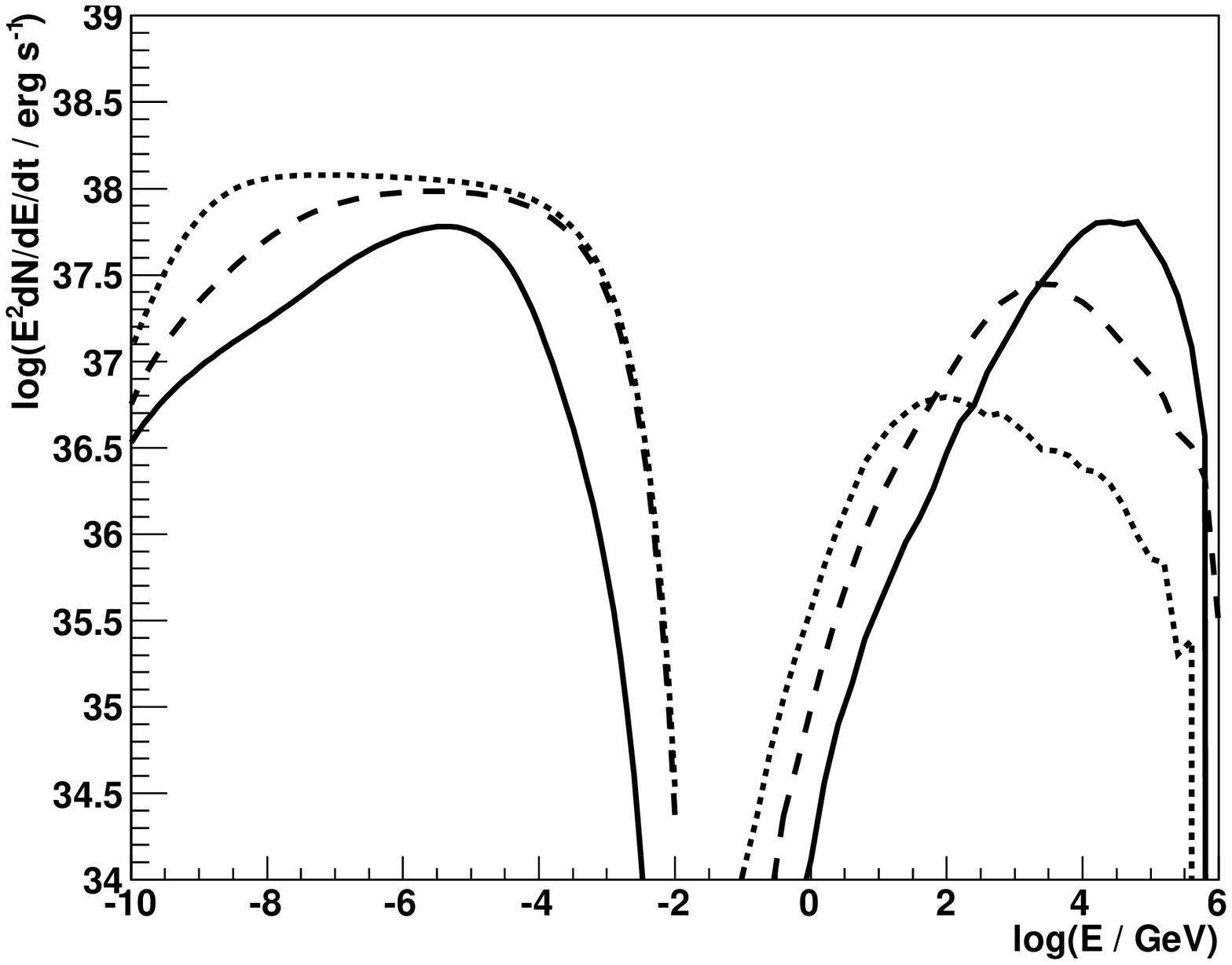}
\caption{As in Fig.~3 but for the case of GCs which entered the jet at the range of distances, 
$l_{\rm min} = 1$ kpc and $l_{\rm max} = 20$ kpc. The results are shown for the Poynting dominated jet,
for the value of magnetization parameter equal to $\mu = 1$ (dotted curves) and the matter dominated jets for $\mu = 0.1$ (dashed), and $\mu = 0.01$ (solid). The spectra are calculated for the acceleration parameter of electrons equal to $\xi = 10^{-3}$ (upper panel) and for $\xi = 10^{-2}$ (bottom panel). The other parameters of the model are these same as in Fig.~3.}
\label{fig4}
\end{figure}
\section{Application to Centaurus A}

Centaurus A is the closest radio galaxy (distance $3.8\pm 0.1$ Mpc) associated with the elliptical galaxy NGC 5128 
(Harris et al.~2010). The SMBH in Cen A has the mass estimated on $M_{\rm BH} = (5.5\pm 3.0)\times 10^7$ M$_\odot$
 (Cappellari et al. 2009). Two jets, clearly asymmetric on kpc scale, are observed in Cen A. The viewing
 angle of the jet has been estimated on $\sim$50$^{\rm o}$ (Tingay et al.~1998, Hardcastle et al. 2003). The projected 
speed of the jet at hundred pc distance scale has been measured on 0.5c (Hardcastle et al. 2003). The power of 
the jet has been estimated on $P_{\rm j}\sim 10^{43}$ erg s$^{-1}$ (Wykes et al.~2013).    

Due to its proximity, X-ray emission is clearly seen from the kpc scale jet in Cen A (Feigelson et al.~1981). 
At the distance between $\sim$1-3 kpcs, the {\it Chandra} observations (0.4-2.5 keV energy range) show that 
the diffusive X-ray emission dominates over X-ray emission from the knots.
This X-ray emission has been interpreted as due to the synchrotron process (Hardcastle et al.~2006). 
Such synchrotron radiation requires the presence of the TeV 
electrons in the kpc scale jet. They have to be accelerated close to the radiation site due to short energy loss time
scales. Recently, a point like $\gamma$-ray source with the power low spectrum (spectral index $\sim$2.7) extending to 
$\sim$2-3 GeV has been detected by {\it Fermi} satellite from the core of Cen A (Abdo et al.~2010a). The analysis of the 4 years of the {\it Fermi} data shows that $\gamma$-ray emission extends up to $\sim$ 50 GeV. However, the spectrum flattens above $\sim$4 GeV (Sahakyan et al.~2013). The spectral index changes from $2.74\pm 0.03$ below 4 GeV to 
$2.09\pm 0.20$ at higher energies. This flattening of the $\gamma$-ray spectrum is consistent with the detection of
the TeV source towards Cen A by the HESS Collaboration (Aharonian et al.~2009). It suggests that the TeV and hard GeV emission might originate in this same mechanism. The lower energy points of these TeV
observations ($>250$ GeV) links nicely to the hard high energy component detected by {\it Fermi} but the spectral
index of the TeV emission is again better described by a steep spectrum (spectral index 
$2.73\pm 0.45_{\rm stat}\pm 0.2_{\rm syst}$, see Aharonian et al.~2009). Such curious spectral behavior, lack of
detected clear variability, and pure angular resolution of $\gamma$-ray observations suggest involvement of at least two radiation processes or emission regions (Sahakyan et al.~2013).   
Due to the large angular extend of the Cen A jets on the sky (of the order of $\sim$10$^{\rm o}$), the {\it Fermi}
Observatory was also able to discover GeV $\gamma$-ray  emission from the regions of the giant radio lobes formed in collisions of jets with intergalactic medium (Abdo et al.~2010b). All these high energy observations show that the
high energy processes are characteristic not only for the direct vicinity of the SMBH but also for the large scale jets. We wonder whether the highest energy component (at $\sim$1 TeV), observed from Cen A by 
the HESS Collaboration, can originate in the kpc scale jet due to the acceleration of electrons on the shocks 
in the jet which are produced in the collisions of compact objects (stars, GCs) with the jet plasma.

In order to test this hypothesis, we performed calculations of the synchrotron and the IC spectra expected in terms of the above jet/compact objects collision model, applying the known parameters of the jet and the surrounding medium of Cen A. The results of calculations are compared with the observed high energy spectrum from Cen A (Fig.~5). In this figure the X-ray spectrum represents 
emission from the kpc scale jet (Hardcastle et al.~2006). The {\it Fermi} spectrum comes from the central part of the galaxy harboring Cen A (Sahakyan et al.~2013). The origin and location of this steady spectral component is at present unknown due to the pure angular resolution of the {\it Fermi}-LAT telescope. Finally, the data points show the measurements of the TeV $\gamma$-ray emission from Cen A inner/intermediate scale jet.
We assumed that the red giant stars and massive stars enter the jet uniformly at distances between 10 pc and 1 kpc from the base of the jet. Electrons are accelerated at the shocks with the energy conversion efficiency of $\eta = 10\%$. 
The spectra are calculated for the presence of $10^6$ red giants (dotted curves) and 200 massive stars (dashed curves) within the jet. The example parameters of these stars are described in Sect.~5.1. 
For the considered parameters of the acceleration  model ($\mu = 0.01$, $\xi = 10^{-3}$), the emission from electrons accelerated on the shock around red giants is not expected to contribute significantly to the high energy spectrum observed from Cen A. On the other hand, about a hundred of massive stars in the jet can produce synchrotron emission 
on the level observed from the jet of Cen A. The accompanying $\gamma$-ray emission, from the componization of background radiation by these same electrons, is on the level of the TeV $\gamma$-ray emission reported by HESS 
from Cen A. We also show the synchrotron and IC spectra produced in collisions of 20 GCs with
the jet at larger distance scale (dot-dashed curves) for these same parameters of the jet. The synchrotron emission is not expected to contribute significantly to the observed X-ray spectrum. However, the IC $\gamma$-ray emission peaks at the TeV energies. The cumulative $\gamma$-ray spectrum from collisions of massive stars and GCs with the jet is clearly consistent with the level of the TeV $\gamma$-ray emission observed by HESS. Therefore, such model might be responsible for the highest energy $\gamma$-ray component observed from Cen A. If the jet in Cen A still moves with the velocity of the order of $\sim0.5c$ over kpc distances, then the calculated spectra (in Fig.~5) should be enhanced by a factor of 
$D^4\sim 2.7$. As a result, our estimates of the number of compact objects within the jet should be reduced by this factor.

We conclude that the steady TeV $\gamma$-ray emission, observed from the central part of the radio galaxy Cen A, can originate in the kpc scale jet. This emission (or at least a part of this emission) should be steady independently on the activity state of the inner jet. Since the expected emission is above the level of sensitivity of the future Cherenkov Telescope Array (CTA), the question on the existence of the low level
persistent emission from the intermediate scale jet in Cen A should be definitively answered in the coming years.

\begin{figure}
\vskip 5.5truecm
\includegraphics{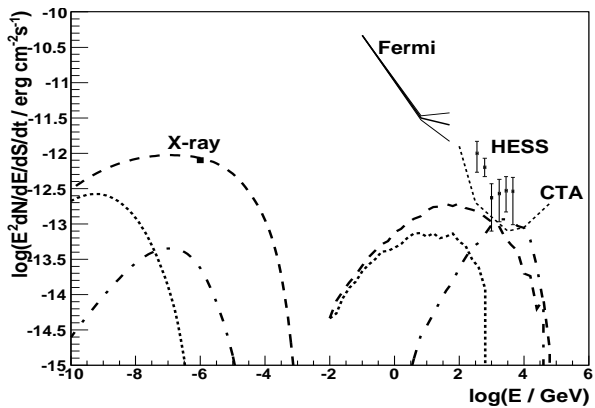}
\caption{The comparison of the high energy observations of Cen A jet with the calculations of the synchrotron and IC spectra expected from the collisions of GCs 
and stars with the jet in Cen A. The X-ray emission from the kpc scale jet in Cen A is taken from Hardcastle et al.~(2006), the {\it Fermi} $\gamma$-ray spectral measurements are from Sahakyan et al.~(2013) and the TeV $\gamma$-ray spectrum measured by HESS from Aharonian et al.~(2009). The emission expected from the interaction of $10^6$ red giants with the jet (dotted curves) and $200$ massive stars (dashed curves) is calculated assuming that these stars enter the jet on the distance scale between 10 pc and 1 kpc. The emission from 20 GCs entering the jet on the distance scale between 1 kpc and 20 kpc (dot-dashed curves). The energy conversion efficiency from the shock to leptons is assumed to be $10\%$. The sensitivity of the CTA in marked by the thin dotted curve.}
\label{fig5}
\end{figure}
\section{Conclusions}

We propose that the compact object/jet collision model can be responsible for the non-thermal emission from the intermediate scale jets observed in close radio galaxies. We consider collisions of different type of objects, starting from the red giant stars from the galactic bulge, through the massive stars from the nuclear central cluster and finishing on the globular clusters in the galactic halo. Collisions produce multiple shocks in the jet on which leptons can be accelerated to TeV energies. These leptons are responsible for the diffusive synchrotron and inverse Compton radiation from the kpc scale jets. Contrary to the inner jet collision models, leptons accelerated on the shocks
do not lose energy close to the shock region (near the stars) but they are injected into the jet. They are transported with the jet plasma to farther distances from the base of the jet losing energy on the mentioned above radiation processes. Therefore, this high energy emission is expected to be steady during the time scales of the jet dimension (i.e. thousands of years), in contrast to the emission expected from collisions of stars with the inner jets
(below $\sim$1 pc, e.g. Bednarek \& Protheroe~1997). Moreover, due to the saturation of the acceleration process of 
leptons by the synchrotron or the advection time scales, the larger energy leptons are expected to be accelerated at farther along the jet. Therefore, non-thermal emission from jets should show clear dependence on the energy with the lower energy $\gamma$-rays produced closer to the base of the jet. We predict that the multi-TeV $\gamma$-rays should have tendency to be produced farther from the base of the jet. In contrast, such feature may not be observed in the case of the synchrotron X-ray emission if the acceleration of leptons is saturated by synchrotron energy losses. 
Only when the maximum electron energies are constrained by their escape (advection) from the acceleration region,
the maximum energies of synchrotron radiation should show energy dependence on the distance from the base of the jet.  

We compared the example spectra, obtained in terms of such a model, with the observations of the intermediate scale jet in Cen A (Fig.~5). It is found that in general the observed level of the X-ray emission can be explained as a result of the synchrotron radiation produced by electrons accelerated on the shocks around massive stars. On the other hand, the TeV $\gamma$-ray emission is expected to be produced by leptons accelerated on the shocks around massive stars (below $\sim$1 TeV) and by electrons accelerated on the shocks around GCs (above $\sim$1 TeV).  
This hypothesis could be tested in the near future with the operation of the next generation $\gamma$-ray telescopes such as CTA. CTA is expected to reach $\sim$1 arc min resolution at 10 TeV and the integral sensitivity of $\sim 2\times 10^{-14}$ TeV cm$^{-2}$ s$^{-1}$ at a few TeV (Acharya et al.~2013).  With such angular resolution 
the TeV $\gamma$-ray emission from the kpc scale jet in Cen A could be resolved and the prediction on the energy dependent emission with the distance from the base of the jet could be confirmed or disproved. 

We applied the collision model for the radio galaxies whose intermediate jets are expected to be rather slow and observed at 
a relatively large angle to the line of sight.
If the jets still moves relativistically at the kpc distance scale, then the non-thermal radiation can be significantly Doppler boosted (see e.g. recent paper by Bosch-Ramon~2015). In such case the collisions of different types of compact objects (massive stars, red giants, globular clusters) with the relativistic jets could provide natural explanation for the powerful non-thermal emission from blazars observed at small angle to then line of sight.

We would like to thank the Referee for useful comments. This work is supported by the grant through the Polish NCN No. 2011/01/B/ST9/00411.






\end{document}